\documentclass[10pt]{IEEEtran}
\usepackage[noadjust]{cite}
\usepackage[utf8]{inputenc}
\usepackage{amsmath}
\usepackage{amssymb}
\usepackage{amsthm}
\usepackage{mathtools}
\usepackage{bm}
\usepackage{bbm}
\usepackage{comment}
\usepackage{tabularx,booktabs}
\usepackage[caption=false]{subfig}
\usepackage[capitalize]{cleveref}
\DeclareMathOperator*{\argmin}{arg\,min}
\DeclareMathOperator*{\argmax}{arg\,max}

\DeclareMathOperator{\E}{\mathbb{E}}
\renewcommand{\epsilon}{\varepsilon}
\usepackage{pgfplots}
\pgfplotsset{compat=1.15}
\usepgfplotslibrary{fillbetween}
\usetikzlibrary{patterns,arrows,plotmarks}
\usepgfplotslibrary{groupplots}
\pgfdeclarelayer{background}
\pgfsetlayers{background,main}
\usetikzlibrary{automata,positioning}
\usetikzlibrary{decorations}
\usetikzlibrary{decorations.markings}
\usepackage{dsfont}

\newcommand{\edit}[1]{#1}

\usepackage{glossaries}    
    
\newacronym[plural=MDPs,firstplural=Markov Decision Processes (MDPs)]{mdp}{MDP}{Markov Decision Process}
\newacronym{iot}{IoT}{Internet of Things}
\newacronym{iiot}{IIoT}{Industrial Internet of Things}
\newacronym{fec}{FEC}{Forward Error Correction}
\newacronym{snr}{SNR}{Signal to Noise Ratio}
\newacronym{mec}{MEC}{Mobile Edge Computing}
\newacronym{dqn}{DQN}{Deep Q Network}
\newacronym{relu}{ReLU}{Rectified Linear Unit}
\newacronym{pmf}{pmf}{probability mass function}
\newacronym{pomdp}{POMDP}{Partially Observable Markov Decision Process}
\newacronym{iid}{IID}{Independent and Identically Distributed}
\newacronym{aoi}{AoI}{Age of Information}
\newacronym{aoii}{AoII}{Age of Incorrect Information}
\newacronym{voi}{VoI}{Value of Information}
\newacronym{uoi}{UoI}{Urgency of Information}
\newacronym{marl}{MARL}{Multi-Agent Reinforcement Learning}
\newacronym{paoi}{PAoI}{Peak Age of Information}
\newacronym{qaoi}{QAoI}{Age of Information at Query}
\newacronym{eaoi}{EAoI}{Effective Age of Information}
\newacronym{mse}{MSE}{Mean Square Error}
\newacronym{mmse}{MMSE}{Minimum Mean Square Error}
\newacronym{leo}{LEO}{Low Earth Orbit}
\newacronym{geo}{GEO}{Geosynchronous Earth Orbit}
\newacronym{pec}{PEC}{Packet Erasure Channel}
\newacronym{rl}{RL}{Reinforcement Learning}
\newacronym{drl}{DRL}{Deep Reinforcement Learning}
\newacronym{pdf}{PDF}{Probability Density Function}
\newacronym{cdf}{CDF}{Cumulative Distribution Function}
\newacronym{pq}{PQ}{Permanent Query}
\newacronym{qapa}{QAPA}{Query Arrival Process Aware}
\newacronym{harq}{HARQ}{Hybrid Automated Repeat Request}
\newacronym{bs}{BS}{Base Station}
\newacronym{maf}{MAF}{Maximum Age First}

\pgfplotsset{
tick label style={font={\footnotesize\color{white!15!black}}},
label style={font={\footnotesize\color{white!15!black}}},
legend style={font={\footnotesize\color{white!15!black}}},
}

\usepgfplotslibrary{external}
\def \mapside {0.24\linewidth}

\def \fwidth {0.95\columnwidth}
\def \fheight {0.5\columnwidth}

\def \lfwidth {0.95\columnwidth}
\def \lfheight {0.6\columnwidth}

\def \cmwidth {0.4\linewidth}
\def \cmheight {0.15\linewidth}

\definecolor{color0}{HTML}{FFD700}
\definecolor{color1}{HTML}{FFB14E}
\definecolor{color2}{HTML}{FA8775}
\definecolor{color3}{HTML}{EA5F94}
\definecolor{color4}{HTML}{CD34B5}
\definecolor{color5}{HTML}{9D02D7}
\definecolor{color6}{HTML}{0000FF}

\title{Goal-Oriented Scheduling in Sensor Networks with Application Timing Awareness}

\author{Josefine Holm,~\IEEEmembership{Student Member,~IEEE,} Federico Chiariotti,~\IEEEmembership{Member,~IEEE,} Anders E. Kal\o{}r,~\IEEEmembership{Member,~IEEE,} 
Beatriz Soret,~\IEEEmembership{Senior Member,~IEEE}, Torben Bach Pedersen,~\IEEEmembership{Distinguished Contributor,~IEEE}, and Petar Popovski,~\IEEEmembership{Fellow,~IEEE}%
\thanks{J. Holm (jho@es.aau.dk), F. Chiariotti (corresponding author, chiariot@dei.unipd.it), A. E. Kal\o{}r (aek@es.aau.dk), B. Soret (bsa@es.aau.dk), and P. Popovski (petarp@es.aau.dk) are with the Department of Electronic Systems, Aalborg University, 9220 Aalborg \O{}st, Denmark. T. B. Pedersen (tbp@cs.aau.dk) is with the Department of Computer Science, Aalborg University, 9220 Aalborg \O{}st, Denmark. F. Chiariotti is also with the Department of Information Engineering, University of Padova, 35131 Padua, Italy. A. E. Kal\o{}r is also with the Department of Electrical and Electronic Engineering, The University of Hong Kong, Hong Kong. B. Soret is also with the Telecommunications Research Institute (TELMA), Universidad de M\'alaga, 29071 M\'alaga, Spain  (bsoret@ic.uma.es). This work was supported by the Independent Research Fund Denmark, Grants No. 8022-00284B ``SEMIOTIC'' and No. 1056-00006B, EU Horizon 2020 Grant no. 957345 ``MORE,'' the Villum Investigator Grant ``WATER'' from the Velux Foundation, Denmark, and the SoE Young Researchers grant ``REDIAL,'' as part of the European Union's Italian National Recovery and Resilience Plan (NRRP).}
}

\begin{document}

\maketitle

\begin{abstract}
Taking inspiration from linguistics, the communications theoretical community has recently shown a significant recent interest in \emph{pragmatic}, or goal-oriented, communication. In this paper, we tackle the problem of pragmatic communication with multiple clients with different, and potentially conflicting, objectives. We capture the goal-oriented aspect through the metric of \gls{voi}, which considers the estimation of the remote process as well as the timing constraints. However, the most common definition of \gls{voi} is simply the \gls{mse} of the whole system state, regardless of the relevance for a specific client. Our work aims to overcome this limitation by including different summary statistics, i.e., value functions of the state, for separate clients, and a diversified query process on the client side, expressed through the fact that different applications may request different functions of the process state at different times. A query-aware \gls{drl} solution based on statically defined \gls{voi} can outperform naive approaches by 15-20\%.
\end{abstract}

\glsresetall

\section{Introduction}\label{sec:intro}

Semantic and goal-oriented communications~\cite{popovski2020semantic} aim to go beyond the traditional domain of communication theory towards optimizing communication systems with respect to a specific task or goal. In~\cite{shannon1949mathematical}, Shannon and Weaver talk about the \emph{semantics} and \emph{effectiveness} levels of the communication problem: semantic communication corresponds to the transmission of the most meaningful information for the given context, while effective or goal-oriented communication aims at satisfying the specific requests of the receiver. Following the nomenclature in linguistics, we denote the latter as \emph{pragmatic} communication. In this sense, pragmatic communication is about transmitting the most relevant information for the receiver in order to attain a certain goal, taking into account both timing constraints and the shared context, which acts as an implicit information channel between the transmitter and receiver.

This new perspective is crucial in the context of the Industry 4.0 and \gls{iiot} paradigms, which aim at automating manufacturing and industrial processes in a flexible and easily reconfigurable fashion. A representative scenario is the remote estimation of the state of a system by a distributed network of low-power sensors, a classical \gls{iot} problem~\cite{soderlund2019optimization}. The recently proposed \gls{voi} metric~\cite{yates2020agesurvey} represents a theoretical tool to model the pragmatics of the monitoring application, as it can seamlessly integrate the timing performance of the communication network with the underlying estimation~\cite{popovski2022perspective}. However, \gls{voi} is often defined in a static manner: the value function is the \gls{mse} of the system state, and there is an application constantly monitoring the process with a faster pace than the updates.

In many \gls{iot} scenarios, this is not true, as there may be \emph{multiple} clients monitoring the same process through an edge node or gateway. Each client may potentially be interested in a different function of the system state at different times~\cite{chiariotti2022query}.
Different \emph{queries} can then correspond to different functions of the state. A classic query is the sample average among the measured values, but non-linear functions and even order statistics can often be useful in industrial settings. For example, the number of state components that are within their normal operation parameters, or the maximum among the state components, can be helpful to trigger safety conditions and shut down machines or raise a warning to the operators. As another example, the sample variance can be useful when monitoring the strain on different components of a building or a structure, such as a bridge. In this case, the goal of the system is implicitly stated in the queries: instead of minimizing the estimation error on the state of the system as a whole, the objective is to reduce the error on the response to the \emph{specific query} that a client makes. The dynamic nature of these queries, which may arrive at different times from different clients, is crucial in the optimization: a pragmatic transmitter needs to take into account not only the relevance of sensor information, but also which query might arrive next.

\begin{figure}[t!]
    \centering
    \resizebox{.95\columnwidth}{!}{
    \input{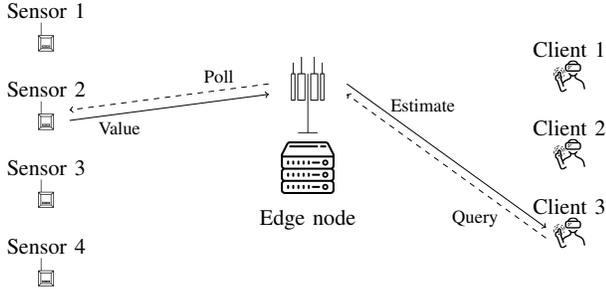}}
    \caption{Illustration of the considered monitoring scenario, where an edge node collects information from sensors to reply to queries from clients. Queries are indicated by dashed lines, while their responses are solid lines.}
    \label{fig:sysmodel}
\end{figure}

The general architecture we propose is shown in Fig.~\ref{fig:sysmodel}. The publish/subscribe model is already the standard for \gls{iot} applications~\cite{hakiri2015publish}, with the edge node acting as a message broker. In the system model in the figure, a \gls{mec}-enabled base station polls a set of sensors, which respond with their latest measurements. The edge node uses the data to estimate the overall state of the process measured by the sensors, and receives queries from client applications, which might be different functions of the state, e.g., the highest value among all sensors, or the number of sensors measuring values in a certain range~\cite{chiariotti2022scheduling}. \edit{In this work, we will consider that the edge node has enough computational power to run the estimation in real-time, but the interplay between the complexity of the scenario and the allocation of computing resources is an interesting problem that could complement our analysis~\cite{du2022sdn}.}

This setup has a clear timing context for defining goal-oriented communication: the edge node should optimize its scheduling strategy to answer queries accurately, taking into account both the nature of the estimation process and the query process. The optimal \gls{voi} scheduling will then consider not only the \emph{accuracy} of the estimate, i.e., the semantic value of the updates, but also \emph{when} the estimation of a particular function of the state will be needed, i.e., the pragmatic relevance of the information to the receiver. In other words, the goal is not static, as in \gls{mse} minimization, but dynamic, as the edge node needs to anticipate upcoming queries and place more importance on updates that will help for those specific functions (e.g., if the edge node expects a client to ask for the highest temperature among all sensors, it should avoid polling sensors that have a high probability of having an extremely low value, even if doing so increases the overall estimation \gls{mse} over the state).

\begin{figure}[t!]
  \centering
\subfloat[%
  VoI-based sensor scheduling. \label{fig:const_err}%
]{\input{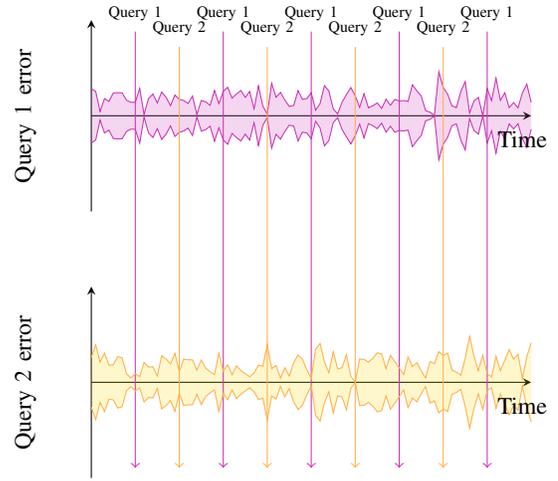}}\\
\subfloat[%
  Pragmatic sensor scheduling. \label{fig:adap_err}%
]{\input{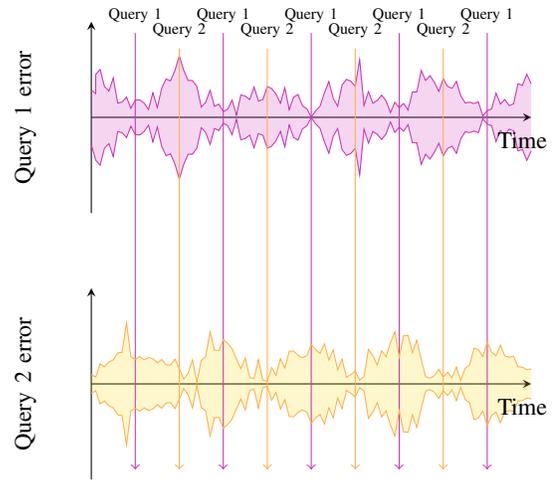}}
    \caption{Sketch of the effects of static \gls{voi}-based and pragmatic scheduling. The lines and colored areas represent the query response confidence interval.}
    \label{fig:err_model}
\end{figure}

Fig.~\ref{fig:err_model} shows a sketch of such an example. In the figure, we plot the instantaneous \gls{mse} over two hypothetical queries. Fig.~\ref{fig:const_err} shows the effect of a \gls{voi} maximization strategy: the errors of both queries are approximately constant across time (as shown by the dashed line), with some fluctuations due to the dynamics of the system and channel errors, independently of the query process. On the other hand, a pragmatic system, as the one shown in Fig.~\ref{fig:adap_err}, can tolerate significant increases in the \gls{mse} of a query as long as the query is not active. Instead, it will aim to reduce the \gls{mse} of the next upcoming query, so that the error \emph{at the time of the query} is minimized. While we expect some fluctuations due to the stochastic nature of these systems, the trend represented by the dashed lines is clear, and leads on average to better responses to specific queries, even if the error when measured at a random time instant is higher. Our previous work~\cite{chiariotti2022query} devised a similar metric for \gls{aoi}, which can be measured at the arrival of a query instead of at any time. While the example is only an illustration of the ideal behavior of a query-aware pragmatic system, our results show that this behavior can indeed be obtained in practice.

In this work, we formulate a pragmatic\footnote{In the remainder of this paper, we use the terms \emph{pragmatic} and \emph{goal-oriented} interchangeably.} scheduling problem, considering both the process and the queries from multiple client applications. We define the problem as a \gls{mdp}, assuming that the edge node has the full statistical knowledge of the process, and solve it with \gls{drl}, as the continuous state space prevents us from using simpler methods such as tabular Q-learning.
However, the main goal of this work is not just to optimize in the specific setting we consider, but rather to provide insights on the novel problem of pragmatic communication, proposing a dynamic definition of \gls{voi} and using a relatively simple use case to understand it better. The contributions of the paper are the following:
\begin{itemize}
 \item We rigorously model the pragmatic sensor scheduling problem with multiple clients, different query functions, and different query generation processes, resulting in a dynamic objective function;
 \item We find a closed-form solution in a simple illustrative example, showing that the optimal policies for different queries can be extremely different and that using the wrong one can lead to significant performance loss;
 \item We frame the problem as a \gls{pomdp}, defining the state and observation space, the reward, and the actions available to the scheduler, and use \gls{drl} to obtain a scheduling solution, which is executable in real time even on an embedded edge processor;
 \item We provide simulation results and comparisons with static \gls{voi} policies and the traditional \gls{maf} policy, which minimizes the average \gls{aoi} of the system, for a scenario in which two query types are present, i.e., a maximum query asking the maximum value among the state components and a count range query asking how many state components have a value within a given interval. The performance of the system is strongly dependent on the queries that arrive to it, as the error on the queries depends on the exact function of the state that they request.
\end{itemize}

The rest of the paper is organized as follows: first, we discuss the state of the art on the topic in Sec.~\ref{sec:related}. We then define the scheduling problem in Sec.~\ref{sec:model}, and model it as an \gls{mdp} in Sec.~\ref{sec:mdp}, which also describes the proposed \gls{drl} solution. Then, Sec.~\ref{sec:sim} describes the simulations comparing the proposed scheme to state-of-the-art policies and their results. Finally, Sec.~\ref{sec:conc} concludes the paper and presents some possible avenues for future work.

\section{Related Work}\label{sec:related}

The problem of designing communication systems that communicate not only reliably, but also effectively and in a timely fashion, has recently received a significant amount of attention. During the last decade, \gls{aoi}~\cite{kaul2012real} and the associated metrics have received a significant attention in the communication engineering community. Most works in the literature study the average \gls{aoi} in queues and networks of queues, using basic queuing theory to compute analytical performance curves~\cite{yates2020agesurvey}.
However, the original definition of the \gls{aoi} can result in suboptimal outcomes if some conditions are not met, and similar metrics that can extend the definition to more general scenarios and applications have been defined~\cite{uysal2022semantic}.

Firstly, \gls{aoi} does not need to be linear: some recent works generalized the definition to measure any non-decreasing function of the age~\cite{sun2019sampling,kosta2020cost}, leading to different considerations in \gls{aoi} optimization, as higher ages are penalized more or less depending on whether the aging function is super- or sub-linear. This can be further generalized by taking into account the actual value of the process tracked by the updates: since \gls{aoi} is a proxy metric for the evolution of the tracking error over time, considering that error directly leads to better performance.  The \gls{aoii}~\cite{maatouk2020AoII} mixes a linear timing penalty with a multiplier based on the error in a discrete system, while \gls{voi}~\cite{ayan2019age} directly measures the error (either real or expected) on the estimation process, whether over a continuous-valued variable or a Markov source~\cite{pappas2021goal}.

The adaptive scheduling of sensors in \gls{iot} scenarios with the goal of minimizing the \gls{aoi} or related metrics is a well-studied problem in the literature. The scheduling problem can be formulated both for multiple sources, in which case it involves balancing the ages of the different sources while avoiding interference~\cite{talak2019optimizing}, or for a single source with resource constraints: usually, these constraints are in the form of limited energy availability~\edit{\cite{moltafet2021power}} or enforced duty cycles\edit{~\cite{emara2020spatiotemporal}. The inclusion of constraints on energy is often combined with energy-harvesting capabilities~\cite{hatami2022demand}.} The most interesting works in this sense consider the \gls{voi}, measured as the expected tracking error of a Kalman filter~\cite{gupta2006stochastic,hashemi2020randomized}. In general, the constraint on the accuracy is due to a communication bottleneck, which occurs due to limited bandwidth and energy, such that sensors need to reduce their transmissions as much as possible~\edit{\cite{nayak2023decentralized}}. The selection of the subset of sensors that will transmit has been studied in this context~\cite{bharti2019value}, often considering the correlation between neighboring sensors' measurements~\cite{zancanaro2023modeling}. Other scenarios in which \gls{voi} is used are data muling applications~\edit{\cite{bui2023scheduling}}, in which drones, robots, or underwater vehicles need to physically move close to sensors to collect the information~\cite{duan2020voi}, and sensor placement problems, in which the issue is not to schedule transmissions, but rather to design the network to maximize accuracy and minimize cost~\cite{hoseyni2019voi}. Our own previous work~\cite{chiariotti2022scheduling} extends the definition of \gls{voi} from the \gls{mse} of the state to arbitrary functions, presenting a one-step optimal scheduling procedure. Another interesting development involves the modeling of the state of each sensor as a Markov chain, posing the polling problem as a \gls{pomdp}~\cite{stamatakis2021autonomous} to identify sensors reporting abnormal values with the minimum energy expenditure~\cite{stamatakis2022semantics}. For a more thorough review of the recent literature on \gls{voi}, we refer the reader to~\cite{alawad2022value}. \edit{Semantic communication is a subject that is closely related to \gls{voi}, and has seen significant developments in the past few years: instead of a pre-defined value, information is evaluated and encoded based on \emph{meaning}, which can only be derived implicitly, either from performance at a given task~\cite{pase2022rate,gunduz2022beyond} or by learning complex patterns in high-dimensional signals such as speech~\cite{weng2023deep} or video~\cite{xia2023wiservr}.}

\begin{table*}[t!]
\centering
	\caption{Main notation used in the system model}
	\label{tab:notation}
    \scriptsize
	\begin{tabular}[c]{ccl|ccl}
		\toprule
		Symbol & Dimension & Meaning & Symbol & Dimension & Meaning\\
		\midrule
        $N$ & 1 & Number of sensors & $M$ & 1 & State dimension\\
        $\mathbf{x}(t)$ & $M\times1$ & System state &  $\mathbf{A}$ & $M\times M$ & State transition matrix\\
        $\mathbf{v}(t)$ & $M\times1$ &  Process noise &$\bm{\Sigma}_v$ &$M\times M$ & Process noise covariance\\
        $\mathbf{y}(t)$ & $N\times 1$ & Observation & $\mathbf{H}$ & $N\times M$ & Observation matrix\\
        $\mathbf{w}(t)$ & $N\times1$ & Observation noise & $\bm{\Sigma}_w$ &$N\times N$ & Observation noise covariance\\
        $\hat{\mathbf{x}}(t)$ & $M\times1$ & Estimated state & $\bm{\psi}(t)$ & $M\times M$ & Estimate error covariance\\
        $\hat{\mathbf{x}}_{\text{pri}}(t)$ & $M\times1$ & \emph{A priori} estimated state & $\bm{\psi}_{\text{pri}}(t)$ & $M\times M$ & \emph{A priori} estimated error covariance\\
        $\varepsilon_n$ & 1 & Packet error probability for sensor $n$ &       $\mathbf{h}(t)$ & $1\times M$ & Observation effect \\
        $\mathds{1}_{a(t)}$ &$1\times N$ & Sensor selection vector & $y_{a(t)}(t)$ & 1 & Received observation\\
        $\mathbf{k}(t)$ & $1\times M$ & Kalman gain & $\sigma_w^2$ & 1 & Received observation variance\\
        $\lambda(t)$ & 1 & Reception indicator variable & $C$ & 1 & Number of clients\\
        $\mathcal{Q}_c$ & 1 & Markov states of client $c$ & $\tilde{}\mathcal{Q}_c$ & 1 & Query states of client $c$\\
        $\mathbf{T}_c$ & $|\mathcal{Q}_c|\times|\mathcal{Q}_c|$ & Transition matrix for client $c$ & $z(\mathbf{x}(t))$ & 1 & Query function\\
        $\hat{z}(\hat{\mathbf{x}}(t))$ & 1 & Query response & $\text{MSE}_z$ & 1 & MSE for query $z$\\
        $S$ & 1 & Number of Monte Carlo samples & $\theta_{c,n}(t)$ & 1 & VoI for sensor $n$ and client $c$\\
        $\mathcal{A}$ & 1  & Scheduler action space & $\mathcal{S}$ & $M^2+M+C$ & Scheduler state space \\
        $\mathcal{O}$ & $M^2+M+C$ & Scheduler observation space & $\pi$ & $\mathcal{O}\rightarrow\Phi(\mathcal{A})$ & Scheduler policy \\ $R(\pi)$ & 1 & Long-term reward function & $\gamma$ & 1 & Discount factor\\
		\bottomrule
	\end{tabular}
\end{table*}

An assumption that \edit{is shared by most of the works we discussed above} is that information is always relevant, and that the application that tracks the process is always active. We can relax this assumption by considering a \emph{query process}, as we did in our previous work~\cite{chiariotti2022query}, in which we defined the \gls{qaoi}: this metric is a sub-sampling of the \gls{aoi}, only considering the instants when a query arrives from the application to be relevant for the optimization. A similar approach was adopted in defining the \gls{eaoi}~\cite{yin2019only}, with a slightly different set of assumptions, and our theoretical results were extended in~\cite{ildiz2022inequality,ildiz2022query}, which showed the different outcome of the \gls{aoi} and \gls{qaoi} minimization problems under an update-or-wait model. Another work also models requests in the optimization function~\cite{hatami2020age}, but it only deals with memoryless request processes, which (as we will describe in the introduction) lead to a solution that is equivalent to standard \gls{aoi} minimization. The extended version of that paper~\cite{hatami2021aoi} considers more complex scenarios with partial battery knowledge, but still uses the same memoryless request model. Finally, a recent work by Xu \emph{et al.}~\cite{xu2020aoi} also considers a memoryless request process, but considers a mix between traditional \gls{aoi} and query-aware metrics.
By only tracking the \gls{aoi} when a query arrives from the application, the communication system considers not only the freshness of the received information, but also when it is needed: if, for example, the application works over discrete time intervals, transmitting more data close to the next query can reduce the bandwidth and energy usage, while maintaining the same or better accuracy from the application's point of view. We note that query awareness, and query prediction in particular, has been considered in the database literature~\cite{bowman2005optimization,ramachandra2012holistic}, and is related to the problem of predicting tasks in edge computing scenarios~\cite{yang20cachenoma,zhang2017mobile}. However, contrary to our work, these do not consider the significance of the fetched data.

The problem of value-oriented scheduling has also been approached in distributed control: if the agents have communication capabilities, the most valuable piece of information is the one that will allow them to improve their performance in the task. The \gls{uoi} metric~\cite{zheng2020urgency} directly considers how much an update would affect a known linear controller. If we consider \gls{marl} agents, the problem is more complex \cite{zhang2013coordinating}, as the communication policy is implicitly learned by the agents while they converge to the optimal control policy, and this approach has only been successful in simple problems~\cite{foerster2016learning} or with only one supporting agent communicating to a primary one~\cite{tung2021effective}. However, because of the centralized nature of our pull-based setting, we limit our focus to the single-agent case, and refer the reader to \cite{oroojlooyjadid2019review} for a review of the cooperative \gls{marl} literature.

This work combines and extends some of the ideas on \gls{voi} sensor scheduling and query awareness, as well as concepts from the semantic communications literature, by considering a system with multiple queries arriving at different times, each of which requires different information on the state of the tracked process, represented by a different \gls{voi} function. To the best of our knowledge, this is the first work to consider this complete system model, and a significant step forward towards full-fledged goal-oriented communications.

\section{System Model}\label{sec:model}

We consider a system in which an edge node receives information from a set of $N$ sensors, indexed by $n\in\{1,2,\ldots,N\}$, and has to respond to \emph{queries} from a set $\mathcal{C}$ of remote clients, with $|\mathcal{C}|=C$. Time is divided into slots, denoted as $t=0,1,2,\ldots$, and in each slot, the edge node can send a request to one sensor, and respond to queries from any number of clients.
In turn, the sensors observe a linear dynamic system, whose state is denoted as $\mathbf{x}(t)\in\mathbb{R}^{M\times1}$. The dimensionality of the process state is $M$, which can be different from the number of sensors $N$ in the general case. The system evolves according to a (potentially time-varying) transition matrix $\mathbf{A}$, with an overlaid error perturbation modeled as a multivariate Gaussian noise:
\begin{equation}
    \mathbf{x}(t)=\mathbf{A}\mathbf{x}(t-1)+\mathbf{v}(t),
\end{equation}
where $\mathbf{v}(t)\sim\mathcal{N}(\mathbf{0}_M,\bm{\Sigma}_v)$. The noise $\mathbf{v}(t)$ is zero-mean, and its covariance matrix is $\bm{\Sigma}_v\in\mathbb{R}^{M\times M}$. The sensors then measure a vector $\mathbf{y}(t)\in\mathbb{R}^{N\times1}$, which represents a linear observation of the state of the system with an added Gaussian measurement noise. We define an observation matrix $\mathbf{H}\in\mathbb{R}^{N\times M}$, which defines the linear function of the state that each function observes:
\begin{equation}
    \mathbf{y}(t)=\mathbf{H}\mathbf{x}(t)+\mathbf{w}(t),\label{eq:obs_full}
\end{equation}
where $\mathbf{w}(t)\sim\mathcal{N}(\mathbf{0}_N,\bm{\Sigma}_w)$. The observation noise $\mathbf{w}(t)$ is also zero-mean, with a covariance matrix $\bm{\Sigma}_w\in\mathbb{R}^{N\times N}$. The main symbols in the model are listed in Table~\ref{tab:notation}, along with their dimensionality and meaning.

\subsection{Remote Kalman Tracking}\label{ssec:kalman}

As the edge node does not know the real state $\mathbf{x}(t)$ of the monitored process, it needs to estimate it. In this work, we use the well-known Kalman filter~\cite{kalman1960new}, which is the optimal solution for linear dynamic systems. We assume that the edge node knows the matrices $\mathbf{A}$, $\mathbf{H}$, $\bm{\Sigma}_v$, and $\bm{\Sigma}_w$, and define vector $\hat{\mathbf{x}}(t)\in\mathbb{R}^{M\times1}$ as the best estimate of the state available to the edge node. The Kalman filter also outputs a covariance matrix $\bm{\psi}(t)$, which corresponds to the expected value $\E[(\mathbf{x}(t)-\hat{\mathbf{x}}(t))^T(\mathbf{x}(t)-\hat{\mathbf{x}}(t))]$.
We can then provide a recursive formula for updating the \emph{a priori} estimate:
\begin{align}
    \hat{\mathbf{x}}_{\text{pri}}(t)=&\mathbf{A}\hat{\mathbf{x}}(t-1)\\
    \bm{\psi}_{\text{pri}}(t)=&\mathbf{A}\bm{\psi}(t-1)\mathbf{A}^T+\bm{\Sigma}_v,
\end{align}
where the $_\text{pri}$ subscript in the estimates indicates that they are \emph{a priori}.

As we stated above, the edge node can request the current value $y_n(t)$ from one sensor per timeslot, whose index is denoted by $a(t)$. We also consider communication errors, modeling the channel between sensor $n$ and the edge node as a \gls{pec} with error probability $\epsilon_n$.  Considering the row vector $\mathbf{h}(t)\in\mathbb{R}^{1\times M}$:
\begin{equation}
    \mathbf{h}(t)=\mathds{1}_{a(t)}\mathbf{H},
\end{equation}
where $\mathds{1}_{a(t)}$ is the row vector of length $N$ whose elements are all 0, except for the one with index $a(t)$, which is equal to 1. We can then update the observation function in~\eqref{eq:obs_full}, getting the value $y_{a(t)}(t)$, which is the observation transmitted by the polled sensor $a(t)$:
\begin{equation}
    y_{a(t)}(t)=\mathbf{h}(t)\mathbf{x}(t)+\mathds{1}_{a(t)}\mathbf{w}(t)=\mathds{1}_{a(t)}\mathbf{y}(t).
\end{equation}
We indicate the outcome of the transmission at time $t$ by the Bernoulli random variable $\lambda(t)$, which is equal to 1 if the transmission is successful and 0 otherwise. In the former case, the edge node receives observation $y_{a(t)}(t)$, while in the latter, it receives no observation for this time step.
We can then give the Kalman gain row vector $\mathbf{k}(t)\in\mathbb{R}^{1\times M}$ as:
\begin{equation}
    \mathbf{k}(t)=\bm{\psi}_{\text{pri}}(t)\mathbf{h}(t)^T\left(\mathbf{h}(t)\bm{\psi}_{\text{pri}}(t)\mathbf{h}(t)^T+\sigma_w^2(t)\right)^{-1},
\end{equation}
where $\sigma_w^2(t)=\mathds{1}_{a(t)}\bm{\Sigma}_w\mathds{1}_{a(t)}^T$.
The update from the \emph{a priori} estimate of the state to the \emph{a posteriori} one is then given by:
\begin{align}\label{eq:kalmanupdate_x}
       \hat{\mathbf{x}}(t)=&\hat{\mathbf{x}}_{\text{pri}}(t)+\lambda(t)\mathbf{k}(t)\left(y_{a(t)}(t)-\mathbf{h}(t)\hat{\mathbf{x}}_{\text{pri}}(t)\right)\\
    \bm{\psi}(t)=&\left(\mathbf{I}_M-\lambda(t)\mathbf{k}(t)\mathbf{h}(t)\right)\bm{\psi}_{\text{pri}}(t),\label{eq:kalmanupdate_psi}
\end{align}
where $\mathbf{I}_M$ is the $M\times M$ identity matrix. We highlight that the \emph{a priori} and \emph{a posteriori} estimates are the same if $\lambda(t)=0$, e.g., no observation is received by the edge node~\cite{sinopoli2004kalman}, as the \emph{a priori} estimate is the best estimate that the edge node can obtain with the information it has received.

\subsection{The Query Process}
We consider a \emph{query} to be a request for either the state $\mathbf{x}(t)$ itself, or the value of a function $z(\mathbf{x}(t))$ of it. The edge node receives queries and responds with an estimate $\hat{z}(\hat{\mathbf{x}}(t),\bm{\psi}(t))$ based on the current state of the Kalman filter.

The temporal element of the query process can be modeled as a Markov chain. We assume that each client $c$ follows an independent Markov chain, whose state at time $t$ is $q_c(t)\in\mathcal{Q}_c$, with a known transition matrix $\mathbf{T}_c$. Each client $c$ always requests the same function $z_c$ anytime its Markov chain is in a subset of states, which we denote as $\tilde{\mathcal{Q}}_c$. Naturally, the state of each client is unknown to the edge node, which can only know which clients are currently subscribed and when did they send their last query. In some cases (e.g., periodic queries), this information is sufficient to predict the next query perfectly, as we will discuss below, but in the general case, the information about the query process available to the edge node entails some randomness and uncertainty.

\subsection{Responding To Queries}\label{ssec:queries}

The objective of the edge node is to respond to queries as accurately as possible, i.e., to minimize the error of its responses. The \gls{mse} for client $c$ is defined as:
\begin{equation}
    \text{MSE}_{z_c}=\mathbb{E}\left[\left(\hat{z}(\hat{\mathbf{x}}(t),\bm{\psi}(t))-z(\mathbf{x}(t))\right)^2\right].
\end{equation}
The edge node can act in two ways to minimize the \gls{mse}: the first is to optimally use its knowledge of the state by using an \gls{mmse} estimator to obtain $\hat{z}(\hat{\mathbf{x}}(t),\bm{\psi}(t))$, and the second is to poll sensors according to the expected \gls{voi} of their readings, i.e., schedule the sensor that can help the most in reducing the \gls{mse} of the next queries. The former problem is relatively simple, while the latter will be tackled in Sec.~\ref{sec:mdp}.

We can give the definitions and \gls{mmse} estimators for some of the most intuitive queries as follows:
\begin{enumerate}
    \item \emph{State}: in this case, the request is for the direct value of $\mathbf{x}(t)$;
    \item \emph{Sample mean}: in this case, the function that the client requests to the edge node is the sample average, represented by:
    \begin{equation}
        z_{\text{avg}}(\mathbf{x}(t))=\frac{1}{M}\sum_{m=1}^M x^{(m)}(t),
    \end{equation}
    where $x^{(m)}(t)$ is the $m$-th element of $\mathbf{x}(t)$;
    \item \emph{Sample variance}: in this case, the sample variance is computed as:
    \begin{equation}
        z_{\text{var}}(\mathbf{x}(t))=\frac{1}{M-1}\sum_{m=1}^M\left(x^{(m)}(t)-z_{\text{avg}}(\mathbf{x}(t))\right)^2;\label{eq:z_var}
    \end{equation}
\item\emph{Maximum}: this query returns the maximum component of the state:
  \begin{equation}
    z_{\text{max}}(\mathbf{x}(t))=\max_{m\in\{1,\ldots,M\}}x^{(m)}(t);
  \end{equation}
    \item\emph{Count range}: this function counts how many components of the state are inside a given interval $[a,b]$. The count range query is defined as:
    \begin{equation}
        z_{\text{cnt}}(\mathbf{x}(t))=\sum_{m=1}^M \mathbbm{1}\left(x^{(m)}(t)\in[a,b]\right),
    \end{equation}
    where $\mathbbm{1}(\cdot)$ is the indicator function, equal to 1 if the condition inside the parentheses is verified and 0 otherwise.
    Note that the definition of the function $z_{\text{cnt}(\cdot)}$ should include the interval $[a,b]$, but here we omit it for the sake of readability. Furthermore, $a$ and $b$ are assumed to be fixed for the same query process.
\end{enumerate}

The one-step optimal scheduler for the first three queries can be computed analytically by finding the choice that minimizes the \gls{mse} of the estimate at the next step, and we derive the one-step optimal schedulers for several queries in our previous work~\cite{chiariotti2022scheduling}. However, the optimal scheduler, and even the \gls{mmse} estimator, for more complex queries like the latter two, with highly non-linear functions, is hard to compute analytically.
In the case of order statistics, a closed-form \gls{mmse} estimator might not even be achievable~\cite{amram1985multivariate}, as the extreme values of high-dimensional multivariate Gaussian variables are computed only as limiting distributions in the relevant literature. In order to compute the \gls{mse} of a given response to the count range query, we have to define the region $\mathcal{Z}(m)$:
\begin{equation}
    \mathcal{Z}(m)=\left\{\mathbf{x}\in\mathbb{R}^{M\times1}: z_{\text{cnt}}(\mathbf{x})=m\right\}.
\end{equation}
We can then define the probability that $z_{\text{cnt}}(\mathbf{x}(t))$ is equal to $m$, corresponding to the integral of the multivariate Gaussian random variable $\mathbf{x}(t)\sim\mathcal{N}(\hat{\mathbf{x}}_t(t),\bm{\psi}(t))$ in $\mathcal{Z}(m)$:
\begin{equation}
  \mathcal{P}(z_{\text{cnt}}(\mathbf{x}(t))=m)=\int\displaylimits_{\mathcal{Z}(m)}\frac{e^{-\frac{1}{2}(\mathbf{x}-\hat{\mathbf{x}}_t(t))^T\bm{\psi}^{-1}(t)(\mathbf{x}-\hat{\mathbf{x}}_t(t))}}{\sqrt{2\pi^M|\bm{\psi}(t)}}  d\mathbf{x}.\label{eq:region_integral}
\end{equation}
The \gls{mmse} estimator for the count range query is then given by:
\begin{equation}
    \hat{z}_{\text{cnt}}(t)=\sum_{m=0}^M m \mathcal{P}(z_{\text{cnt}}(\mathbf{x}(t))=m).
\end{equation}
The corresponding \gls{mse} is defined as follows:
\begin{equation}
    \text{MSE}_{\text{cnt}}(t)=\sum_{m=0}^M (m-\hat{z}_{\text{cnt}}(t))^2\mathcal{P}(z_{\text{cnt}}(\mathbf{x}(t))=m).
\end{equation}
The integral in~\eqref{eq:region_integral} can only be computed numerically, and is extremely hard to tabulate.

We can then consider the \gls{voi}, which is used to evaluate the quality of a scheduler. If a query of type $z_c$ arrives at step $t$, we define the value of the information available to sensor $n$ as the expected reduction in the \gls{mse} for that query with respect to the \emph{a priori} estimate. The \gls{voi} $\theta_{c,n}(t)$ is then given by:
\begin{equation}
\begin{aligned}
 \theta_{c,n}(t)=(1-\varepsilon_n)\mathbb{E}\left[\left(\hat{z}_c(\hat{\mathbf{x}}_{\text{pri}}(t),\bm{\psi}_{\text{pri}}(t))-z_c(\mathbf{x}(t))\right)^2\right]\\
 -(1-\varepsilon_n)\mathbb{E}\left[\left(\hat{z}_c(\hat{\mathbf{x}}(t),\bm{\psi}(t))-z_c(\mathbf{x}(t))\right)^2\big|a_t=n\right].
 \end{aligned}\label{eq:voi}
\end{equation}

The one-step optimal \gls{voi} scheduler for any query function can be easily approximated using Monte Carlo methods~\cite{luengo2020survey} by drawing samples from the \emph{a priori} distribution. The detailed algorithm for Monte Carlo-based scheduling is given in our previous work~\cite{chiariotti2022scheduling}. While Monte Carlo estimates are not \gls{mmse}, they approach the optimal estimator as the number of samples grows to infinity, at the cost of computational complexity. If we consider $S$ samples, the complexity of one Monte Carlo estimate is $O(SM^2)$.

\section{The Scheduling Problem}\label{sec:mdp}
In the previous section, we defined the system model and determined the optimal estimator for common query functions, along with a Monte Carlo strategy for general functions. However, the most complex problem is not to reply directly to a query, but to consider future queries in a foresighted manner, scheduling sensor transmissions so as to minimize the \gls{mse} on future responses. This requires to consider not only the monitored system, but also the query process and the interplay between different query functions. For example, two clients which request the maximum and minimum will need very different parts of the state to be estimated accurately, and balancing between their needs will be complex. The polling decisions made by the edge node also affect the future state of the Kalman filter, requiring a dynamic strategy.

We can model the scheduling problem for the edge node as a \gls{pomdp}, in which the edge node must decide which sensor to poll at each time slot. The action space is then simply $\mathcal{A}=\{1,\ldots,N\}$, while the state space is more complex. The state of the Kalman filter just before the update, described by $\hat{\mathbf{x}}_{\text{pri}}(t)$ and $\bm{\psi}_{\text{pri}}(t)$, is included in the state, and so should all the states of the clients. The state space for a system with $C$ clients is then $\mathcal{S}=\mathbb{R}^{M^2+M}\times\prod_{c=1}^C\mathcal{Q}_c$, as the state is given by the tuple $s(t)=\left(\hat{\mathbf{x}}_{\text{pri}}(t),\bm{\psi}_{\text{pri}}(t),q_1(t),\ldots,q_C(t)\right)$. However, the edge node does not know the state $q_c(t)$ of each client, but only the time that has passed since the last query, which we define as $\tau_c(t)\in\mathbb{N}$. We then have an observation tuple $o(t)=\left(\hat{\mathbf{x}}_{\text{pri}}(t),\bm{\psi}_{\text{pri}}(t),\tau_1(t),\ldots,\tau_C(t)\right)$, belonging to the observation space $\mathcal{O}=\mathbb{R}^{M^2+M}\times\mathbb{N}^C$. The matrices $\mathbf{A}$, $\mathbf{H}$, $\bm{\Sigma}_v$, and $\bm{\Sigma}_w$, as well as the error probability vector $\bm{\epsilon}=[\epsilon_n]$ and the query functions $z_c$, should also be known \emph{a priori} to the edge node, but are not part of the state.

Note that the problem reduces to a fully observable \gls{mdp} if the time since the last transmission is sufficient to determine the next query, i.e., if the following condition is true:
\begin{equation}
\begin{aligned}
    \mathcal{P}(q_c(t+1)\in\tilde{\mathcal{Q}}_c|q_c(t)=q)=\mathcal{P}(q_c(t+1)\in\tilde{\mathcal{Q}}_c|\tau_c(t)),\\
    \forall q\in\mathcal{Q}_c,\tau_c(t)\in\mathbb{N}.
\end{aligned}\label{eq:observable}
\end{equation}
Two special cases of this are the memoryless process, in which the Markov chain only has two states (query and no query), and the deterministic chain with $|\tilde{\mathcal{Q}}_c|=1$, which leads to a periodic query process. In the general case, the state of the query process depends on external factors (e.g., a human operator), and is not directly knowable by the edge node: if a stochastic transition can lead to a state in which~\eqref{eq:observable} is not verified, the problem is partially observable.

The transition probability $P(s,s'|a)$ from one state to the next for a given action is then determined by the Markov chains of each client, along with the Kalman filter equations in~\eqref{eq:kalmanupdate_x} and~\eqref{eq:kalmanupdate_psi}. The final parameter to define the \gls{pomdp} is then the reward function $r(t)$:
\begin{equation}
 r(t)=-\sum_{c\in\mathcal{C}}\alpha_c\text{MSE}_{z_c}(t)\mathbbm{1}(q_c(t)\in\tilde{\mathcal{Q}}_c),\label{eq:reward}
\end{equation}
where $\alpha_c>0$ is a weight parameter representing the relative importance of each client, whose value is given by the system designers, and is thus known \emph{a priori} by the edge node. The reward is always negative, as the objective is to minimize the error on all queries.

We then define a \emph{policy} $\pi:\mathcal{O}\rightarrow\Phi(\mathcal{A})$, where $\Phi(\mathcal{A})$ is a probability distribution over the action space $\mathcal{A}$. In other words, the policy maps observed states to the probability of selecting each sensor. We can then define the long-term reward function  $R(t|\pi)$:
\begin{equation}
 R(\pi)=\E\left[\sum_{t=0}^\infty \gamma^{t}r(t)\Big| s_0,\pi\right],\label{eq:longterm}
\end{equation}
where $\gamma\in[0,1)$ is an exponential discount factor. The objective of the scheduling problem is then to find the optimal policy $\pi^*$, which maximizes the long-term reward:
\begin{equation}
 \pi^*=\argmax_{\pi:\mathcal{O}\rightarrow\Phi(\mathcal{A})}R(\pi).
\end{equation}
The case for $\gamma=0$ is a special case, in which future steps are never counted, and only performance in the next step matters: this case was solved analytically in our previous work~\cite{chiariotti2022scheduling}.

\subsection{A Simple Example: The Effect of Queries on the Optimal Policy}\label{ssec:simple}

We can first consider a simple example, in which a system with $N=2$ sensors observes a process with $M=2$ and needs to reply to a single client (i.e., $C=1$). The communication is assumed to be error-free, and each sensor $m$ observes an independent binary Markov chain with state $\mathbf{x}(t)\in\{0,1\}^M\,\forall t$ and $\mathbf{H}=\mathbf{I}_2$. At each time step, the state changes with probability $p_m$ and remains the same with probability $1-p_m$, so that the transition matrix $\mathbf{T}_m$ is given by:
\begin{equation}
\mathbf{T}_m=\begin{pmatrix}
1-p_m & p_m \\
p_m & 1-p_m
\end{pmatrix}.
\end{equation}
We know that the observation of the state is error-free, so after $\Delta_m$ steps from the last observation $o_m$, the \emph{a posteriori} state probability distribution of Markov chain $m$ is given by:
\begin{equation}
  P_m(\Delta_m,o_m)=\mathbf{T}_m^{\Delta_m}\begin{pmatrix}
1-o_m &
o_m 
\end{pmatrix}^T.
\end{equation}
We assume that a query arrives at every step from the client, but define two types of clients with different query functions: the first client asks for a maximum query, while the second asks for a count query, i.e., how many sensors measure a value of 1.
If at least one of the sensors has a value of 1, the value of the other sensor is useless for the maximum query; on the other hand, it is still relevant for the count query. We can compute the \gls{mmse} response to each query:
\begin{align}
  \hat{z}_{\text{max}}(\bm{\Delta},\mathbf{o})=& 1-P_{1,0}(\Delta_1,o_1)P_{2,0}(\Delta_2,o_2);\\
  \hat{z}_{\text{cnt}}(\bm{\Delta},\mathbf{o})=&P_{1,1}(\Delta_1,o_1)+P_{2,1}(\Delta_2,o_2),
\end{align}
where $P_{m,i}(\Delta_m,o_m)$ is the \emph{a posteriori} probability that chain $m$ will be in state $i$, given the latest observation and its age, and $\bm{\Delta}$ and $\mathbf{o}$ are the vectors of ages and observed values, respectively. We highlight that, aside from a few special cases, the query response will be a value between 0 and 1, corresponding to the probability of the correct response being 1, as this is the \gls{mmse} estimator for Bernoulli random variables. We can also compute the \gls{mse} for both queries:
\begin{align}
  \text{MSE}_{\text{max}}(\bm{\Delta},\mathbf{o})=& P_{1,0}(\Delta_1,o_1)P_{2,0}(\Delta_2,o_2)\\
  &\times(1-P_{1,0}(\Delta_1,o_1)P_{2,0}(\Delta_2,o_2));\\
  \text{MSE}_{\text{cnt}}(\bm{\Delta},\mathbf{o})=&P_{1,0}(\Delta_1,o_1)+P_{2,0}(\Delta_2,o_2)\\&-\left(P_{1,0}(\Delta_1,o_1)\right)^2-\left(P_{2,0}(\Delta_2,o_2)\right)^2.
\end{align}
We can note that, if we observe one of the two states and $o=1$, the response to the maximum query is always correct, as the probability of that component being equal to 0 is 0. In order to maximize the long-term reward from~\eqref{eq:longterm}, we can adopt the classical policy iteration method, as described in \cite[Ch. 4]{sutton2018reinforcement} after truncating the \gls{pomdp} by setting a maximum age $\Delta_{\text{max}}$. While policy iteration is not directly applicable to \glspl{pomdp}, we can recast the problem as a fully observable \gls{mdp} whose states fully describe the history of observations~\cite{hansen1997improved}. In our case, the time since the last observation of each state variable and the values of those observations are enough to enjoy this property. The expanded state is then defined as $s=(\bm{\Delta},\mathbf{o})\in\mathcal{S}$, over which we can apply policy iteration. The transitions from one state to the other are extremely simple, and we can easily derive the transition probability $\mathcal{P}(s(t+1)|s(t),\pi(s(t)))$.

Policy iteration has two steps, called evaluation and improvement. The algorithm is initialized with an approximate value $V_0(s)$ for each state and a policy $\pi_0$, which can be set as all zeros. It then repeats the two steps iteratively until the policy converges. In the first step at iteration $t$, the value function $V_t(s)$ is updated as follows:
\begin{equation}
  V_{t+1}(s)=-\text{MSE}(s,\pi_t(s))+\gamma \sum_{s'\in\mathcal{S}}\mathcal{P}(s'|s,\pi_t(s))V_t(s').
\end{equation}
Naturally, the definition of the \gls{mse} depends on the type of query. After the value has been updated for all states, the policy is updated:
\begin{equation}
  \pi_{t+1}(s)=\argmin_{a\in\{1,2\}} -\text{MSE}(s,a)+\sum_{s'\in\mathcal{S}}\mathcal{P}(s'|s,a)V_{t+1}(s').
\end{equation}
Policy iteration is guaranteed to converge to the optimal policy in finite-state \glspl{mdp} with finite reward~\cite{howard1960dynamic}. We can also note that the formulation corresponds to maximizing the \gls{voi} $\theta_{c,m}(t)$ for the selected query, as defined in~\eqref{eq:voi}.

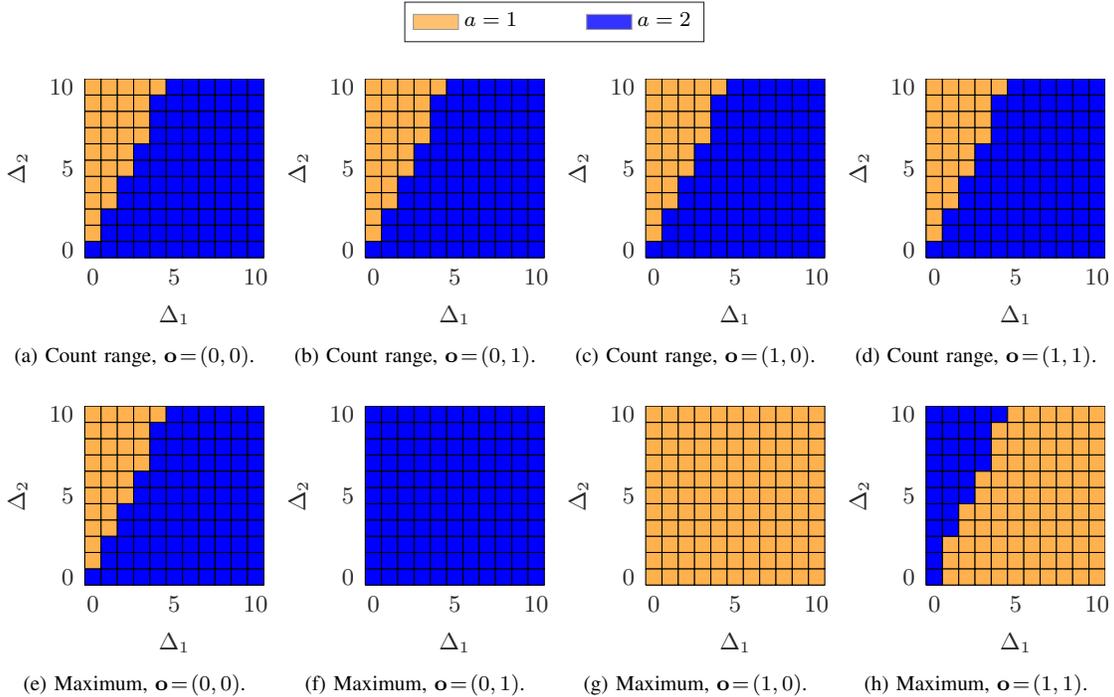
\begin{figure*}[t!]
    \centering    
    \subfloat{
\begin{tikzpicture}

\begin{axis}[
    width=0,
    height=0,
    at={(0,0)},
    scale only axis,
    xmin=0,
    xmax=0,
    xtick={},
    ymin=0,
    ymax=0,
    ytick={},
    axis background/.style={fill=white},
    legend style={legend cell align=center, align=center, draw=white!15!black, font=\scriptsize, at={(0, 0)}, anchor=center, /tikz/every even column/.append style={column sep=2em}},
    legend columns=10,
]
\addplot[area legend,patch type=square,color=gray,fill=color1,opacity=0.8]
table[row sep=crcr, point meta=\thisrow{c}] {%
x   y   c\\
1   0  0\\
0   1  0\\
};
\addlegendentry{$a=1$};
  
\addplot[area legend,patch type=square,color=gray,fill=color6,opacity=0.8]
table[row sep=crcr] {%
0   0\\
1   1\\
};
\addlegendentry{$a=2$};

\end{axis}

\end{tikzpicture}}\\
    \setcounter{subfigure}{0}
\subfloat[Count range, $\mathbf{o}\!=\!(0,0)$.
  \label{fig:cnt00}%
]{
%
%
\begin{tikzpicture}

\begin{axis}[%
width=\mapside,
height=\mapside,
xmin=-0.5,
xmax=10.5,
xlabel={$\Delta_1$},
ylabel={$\Delta_2$},
ymin=-0.5,
ymax=10.5,
xtick
axis background/.style={fill=white},
axis x line*=bottom,
axis y line*=left,
xmajorgrids,
ymajorgrids
]
\addplot[%
surf,
shader=flat corner, draw=black, colormap={mymap}{[1pt] rgb(0pt)=(1,0.694,0.306); rgb(255pt)=(0,0,1)}, mesh/rows=12]
table[row sep=crcr, point meta=\thisrow{c}] {%
x	y	c\\
-0.5	-0.5	2\\
-0.5	0.5	1\\
-0.5	1.5	1\\
-0.5	2.5	1\\
-0.5	3.5	1\\
-0.5	4.5	1\\
-0.5	5.5	1\\
-0.5	6.5	1\\
-0.5	7.5	1\\
-0.5	8.5	1\\
-0.5	9.5	1\\
-0.5	10.5	1\\
0.5	-0.5	2\\
0.5	0.5	2\\
0.5	1.5	2\\
0.5	2.5	1\\
0.5	3.5	1\\
0.5	4.5	1\\
0.5	5.5	1\\
0.5	6.5	1\\
0.5	7.5	1\\
0.5	8.5	1\\
0.5	9.5	1\\
0.5	10.5	1\\
1.5	-0.5	2\\
1.5	0.5	2\\
1.5	1.5	2\\
1.5	2.5	2\\
1.5	3.5	2\\
1.5	4.5	1\\
1.5	5.5	1\\
1.5	6.5	1\\
1.5	7.5	1\\
1.5	8.5	1\\
1.5	9.5	1\\
1.5	10.5	1\\
2.5	-0.5	2\\
2.5	0.5	2\\
2.5	1.5	2\\
2.5	2.5	2\\
2.5	3.5	2\\
2.5	4.5	2\\
2.5	5.5	2\\
2.5	6.5	1\\
2.5	7.5	1\\
2.5	8.5	1\\
2.5	9.5	1\\
2.5	10.5	1\\
3.5	-0.5	2\\
3.5	0.5	2\\
3.5	1.5	2\\
3.5	2.5	2\\
3.5	3.5	2\\
3.5	4.5	2\\
3.5	5.5	2\\
3.5	6.5	2\\
3.5	7.5	2\\
3.5	8.5	2\\
3.5	9.5	1\\
3.5	10.5	1\\
4.5	-0.5	2\\
4.5	0.5	2\\
4.5	1.5	2\\
4.5	2.5	2\\
4.5	3.5	2\\
4.5	4.5	2\\
4.5	5.5	2\\
4.5	6.5	2\\
4.5	7.5	2\\
4.5	8.5	2\\
4.5	9.5	2\\
4.5	10.5	2\\
5.5	-0.5	2\\
5.5	0.5	2\\
5.5	1.5	2\\
5.5	2.5	2\\
5.5	3.5	2\\
5.5	4.5	2\\
5.5	5.5	2\\
5.5	6.5	2\\
5.5	7.5	2\\
5.5	8.5	2\\
5.5	9.5	2\\
5.5	10.5	2\\
6.5	-0.5	2\\
6.5	0.5	2\\
6.5	1.5	2\\
6.5	2.5	2\\
6.5	3.5	2\\
6.5	4.5	2\\
6.5	5.5	2\\
6.5	6.5	2\\
6.5	7.5	2\\
6.5	8.5	2\\
6.5	9.5	2\\
6.5	10.5	2\\
7.5	-0.5	2\\
7.5	0.5	2\\
7.5	1.5	2\\
7.5	2.5	2\\
7.5	3.5	2\\
7.5	4.5	2\\
7.5	5.5	2\\
7.5	6.5	2\\
7.5	7.5	2\\
7.5	8.5	2\\
7.5	9.5	2\\
7.5	10.5	2\\
8.5	-0.5	2\\
8.5	0.5	2\\
8.5	1.5	2\\
8.5	2.5	2\\
8.5	3.5	2\\
8.5	4.5	2\\
8.5	5.5	2\\
8.5	6.5	2\\
8.5	7.5	2\\
8.5	8.5	2\\
8.5	9.5	2\\
8.5	10.5	2\\
9.5	-0.5	2\\
9.5	0.5	2\\
9.5	1.5	2\\
9.5	2.5	2\\
9.5	3.5	2\\
9.5	4.5	2\\
9.5	5.5	2\\
9.5	6.5	2\\
9.5	7.5	2\\
9.5	8.5	2\\
9.5	9.5	2\\
9.5	10.5	2\\
10.5	-0.5	2\\
10.5	0.5	2\\
10.5	1.5	2\\
10.5	2.5	2\\
10.5	3.5	2\\
10.5	4.5	2\\
10.5	5.5	2\\
10.5	6.5	2\\
10.5	7.5	2\\
10.5	8.5	2\\
10.5	9.5	2\\
10.5	10.5	2\\
};
\end{axis}
\end{tikzpicture}%
}
\subfloat[%
  Count range, $\mathbf{o}\!=\!(0,1)$.
  \label{fig:cnt01}%
]{
%
%
\begin{tikzpicture}

\begin{axis}[%
width=\mapside,
height=\mapside,
xmin=-0.5,
xmax=10.5,
xlabel={$\Delta_1$},
ylabel={$\Delta_2$},
ymin=-0.5,
ymax=10.5,
xtick
axis background/.style={fill=white},
axis x line*=bottom,
axis y line*=left,
xmajorgrids,
ymajorgrids
]
\addplot[%
surf,
shader=flat corner, draw=black, colormap={mymap}{[1pt] rgb(0pt)=(1,0.694,0.306);; rgb(255pt)=(0,0,1)}, mesh/rows=12]
table[row sep=crcr, point meta=\thisrow{c}] {%
x	y	c\\
-0.5	-0.5	2\\
-0.5	0.5	1\\
-0.5	1.5	1\\
-0.5	2.5	1\\
-0.5	3.5	1\\
-0.5	4.5	1\\
-0.5	5.5	1\\
-0.5	6.5	1\\
-0.5	7.5	1\\
-0.5	8.5	1\\
-0.5	9.5	1\\
-0.5	10.5	1\\
0.5	-0.5	2\\
0.5	0.5	2\\
0.5	1.5	2\\
0.5	2.5	1\\
0.5	3.5	1\\
0.5	4.5	1\\
0.5	5.5	1\\
0.5	6.5	1\\
0.5	7.5	1\\
0.5	8.5	1\\
0.5	9.5	1\\
0.5	10.5	1\\
1.5	-0.5	2\\
1.5	0.5	2\\
1.5	1.5	2\\
1.5	2.5	2\\
1.5	3.5	2\\
1.5	4.5	1\\
1.5	5.5	1\\
1.5	6.5	1\\
1.5	7.5	1\\
1.5	8.5	1\\
1.5	9.5	1\\
1.5	10.5	1\\
2.5	-0.5	2\\
2.5	0.5	2\\
2.5	1.5	2\\
2.5	2.5	2\\
2.5	3.5	2\\
2.5	4.5	2\\
2.5	5.5	2\\
2.5	6.5	1\\
2.5	7.5	1\\
2.5	8.5	1\\
2.5	9.5	1\\
2.5	10.5	1\\
3.5	-0.5	2\\
3.5	0.5	2\\
3.5	1.5	2\\
3.5	2.5	2\\
3.5	3.5	2\\
3.5	4.5	2\\
3.5	5.5	2\\
3.5	6.5	2\\
3.5	7.5	2\\
3.5	8.5	2\\
3.5	9.5	1\\
3.5	10.5	1\\
4.5	-0.5	2\\
4.5	0.5	2\\
4.5	1.5	2\\
4.5	2.5	2\\
4.5	3.5	2\\
4.5	4.5	2\\
4.5	5.5	2\\
4.5	6.5	2\\
4.5	7.5	2\\
4.5	8.5	2\\
4.5	9.5	2\\
4.5	10.5	2\\
5.5	-0.5	2\\
5.5	0.5	2\\
5.5	1.5	2\\
5.5	2.5	2\\
5.5	3.5	2\\
5.5	4.5	2\\
5.5	5.5	2\\
5.5	6.5	2\\
5.5	7.5	2\\
5.5	8.5	2\\
5.5	9.5	2\\
5.5	10.5	2\\
6.5	-0.5	2\\
6.5	0.5	2\\
6.5	1.5	2\\
6.5	2.5	2\\
6.5	3.5	2\\
6.5	4.5	2\\
6.5	5.5	2\\
6.5	6.5	2\\
6.5	7.5	2\\
6.5	8.5	2\\
6.5	9.5	2\\
6.5	10.5	2\\
7.5	-0.5	2\\
7.5	0.5	2\\
7.5	1.5	2\\
7.5	2.5	2\\
7.5	3.5	2\\
7.5	4.5	2\\
7.5	5.5	2\\
7.5	6.5	2\\
7.5	7.5	2\\
7.5	8.5	2\\
7.5	9.5	2\\
7.5	10.5	2\\
8.5	-0.5	2\\
8.5	0.5	2\\
8.5	1.5	2\\
8.5	2.5	2\\
8.5	3.5	2\\
8.5	4.5	2\\
8.5	5.5	2\\
8.5	6.5	2\\
8.5	7.5	2\\
8.5	8.5	2\\
8.5	9.5	2\\
8.5	10.5	2\\
9.5	-0.5	2\\
9.5	0.5	2\\
9.5	1.5	2\\
9.5	2.5	2\\
9.5	3.5	2\\
9.5	4.5	2\\
9.5	5.5	2\\
9.5	6.5	2\\
9.5	7.5	2\\
9.5	8.5	2\\
9.5	9.5	2\\
9.5	10.5	2\\
10.5	-0.5	2\\
10.5	0.5	2\\
10.5	1.5	2\\
10.5	2.5	2\\
10.5	3.5	2\\
10.5	4.5	2\\
10.5	5.5	2\\
10.5	6.5	2\\
10.5	7.5	2\\
10.5	8.5	2\\
10.5	9.5	2\\
10.5	10.5	2\\
};
\end{axis}
\end{tikzpicture}%
}\subfloat[%
  Count range, $\mathbf{o}\!=\!(1,0)$.
  \label{fig:cnt10}%
]{
%
%
\begin{tikzpicture}

\begin{axis}[%
width=\mapside,
height=\mapside,
xmin=-0.5,
xmax=10.5,
xlabel={$\Delta_1$},
ylabel={$\Delta_2$},
ymin=-0.5,
ymax=10.5,
xtick
axis background/.style={fill=white},
axis x line*=bottom,
axis y line*=left,
xmajorgrids,
ymajorgrids
]
\addplot[%
surf,
shader=flat corner, draw=black, colormap={mymap}{[1pt] rgb(0pt)=(1,0.694,0.306);; rgb(255pt)=(0,0,1)}, mesh/rows=12]
table[row sep=crcr, point meta=\thisrow{c}] {%
x	y	c\\
-0.5	-0.5	2\\
-0.5	0.5	1\\
-0.5	1.5	1\\
-0.5	2.5	1\\
-0.5	3.5	1\\
-0.5	4.5	1\\
-0.5	5.5	1\\
-0.5	6.5	1\\
-0.5	7.5	1\\
-0.5	8.5	1\\
-0.5	9.5	1\\
-0.5	10.5	1\\
0.5	-0.5	2\\
0.5	0.5	2\\
0.5	1.5	2\\
0.5	2.5	1\\
0.5	3.5	1\\
0.5	4.5	1\\
0.5	5.5	1\\
0.5	6.5	1\\
0.5	7.5	1\\
0.5	8.5	1\\
0.5	9.5	1\\
0.5	10.5	1\\
1.5	-0.5	2\\
1.5	0.5	2\\
1.5	1.5	2\\
1.5	2.5	2\\
1.5	3.5	2\\
1.5	4.5	1\\
1.5	5.5	1\\
1.5	6.5	1\\
1.5	7.5	1\\
1.5	8.5	1\\
1.5	9.5	1\\
1.5	10.5	1\\
2.5	-0.5	2\\
2.5	0.5	2\\
2.5	1.5	2\\
2.5	2.5	2\\
2.5	3.5	2\\
2.5	4.5	2\\
2.5	5.5	2\\
2.5	6.5	1\\
2.5	7.5	1\\
2.5	8.5	1\\
2.5	9.5	1\\
2.5	10.5	1\\
3.5	-0.5	2\\
3.5	0.5	2\\
3.5	1.5	2\\
3.5	2.5	2\\
3.5	3.5	2\\
3.5	4.5	2\\
3.5	5.5	2\\
3.5	6.5	2\\
3.5	7.5	2\\
3.5	8.5	2\\
3.5	9.5	1\\
3.5	10.5	1\\
4.5	-0.5	2\\
4.5	0.5	2\\
4.5	1.5	2\\
4.5	2.5	2\\
4.5	3.5	2\\
4.5	4.5	2\\
4.5	5.5	2\\
4.5	6.5	2\\
4.5	7.5	2\\
4.5	8.5	2\\
4.5	9.5	2\\
4.5	10.5	2\\
5.5	-0.5	2\\
5.5	0.5	2\\
5.5	1.5	2\\
5.5	2.5	2\\
5.5	3.5	2\\
5.5	4.5	2\\
5.5	5.5	2\\
5.5	6.5	2\\
5.5	7.5	2\\
5.5	8.5	2\\
5.5	9.5	2\\
5.5	10.5	2\\
6.5	-0.5	2\\
6.5	0.5	2\\
6.5	1.5	2\\
6.5	2.5	2\\
6.5	3.5	2\\
6.5	4.5	2\\
6.5	5.5	2\\
6.5	6.5	2\\
6.5	7.5	2\\
6.5	8.5	2\\
6.5	9.5	2\\
6.5	10.5	2\\
7.5	-0.5	2\\
7.5	0.5	2\\
7.5	1.5	2\\
7.5	2.5	2\\
7.5	3.5	2\\
7.5	4.5	2\\
7.5	5.5	2\\
7.5	6.5	2\\
7.5	7.5	2\\
7.5	8.5	2\\
7.5	9.5	2\\
7.5	10.5	2\\
8.5	-0.5	2\\
8.5	0.5	2\\
8.5	1.5	2\\
8.5	2.5	2\\
8.5	3.5	2\\
8.5	4.5	2\\
8.5	5.5	2\\
8.5	6.5	2\\
8.5	7.5	2\\
8.5	8.5	2\\
8.5	9.5	2\\
8.5	10.5	2\\
9.5	-0.5	2\\
9.5	0.5	2\\
9.5	1.5	2\\
9.5	2.5	2\\
9.5	3.5	2\\
9.5	4.5	2\\
9.5	5.5	2\\
9.5	6.5	2\\
9.5	7.5	2\\
9.5	8.5	2\\
9.5	9.5	2\\
9.5	10.5	2\\
10.5	-0.5	2\\
10.5	0.5	2\\
10.5	1.5	2\\
10.5	2.5	2\\
10.5	3.5	2\\
10.5	4.5	2\\
10.5	5.5	2\\
10.5	6.5	2\\
10.5	7.5	2\\
10.5	8.5	2\\
10.5	9.5	2\\
10.5	10.5	2\\
};
\end{axis}
\end{tikzpicture}%
}\subfloat[%
  Count range, $\mathbf{o}\!=\!(1,1)$.
  \label{fig:cnt11}%
]{
%
%
\begin{tikzpicture}

\begin{axis}[%
width=\mapside,
height=\mapside,
xmin=-0.5,
xmax=10.5,
xlabel={$\Delta_1$},
ylabel={$\Delta_2$},
ymin=-0.5,
ymax=10.5,
xtick
axis background/.style={fill=white},
axis x line*=bottom,
axis y line*=left,
xmajorgrids,
ymajorgrids
]
\addplot[%
surf,
shader=flat corner, draw=black, colormap={mymap}{[1pt] rgb(0pt)=(1,0.694,0.306);; rgb(255pt)=(0,0,1)}, mesh/rows=12]
table[row sep=crcr, point meta=\thisrow{c}] {%
x	y	c\\
-0.5	-0.5	2\\
-0.5	0.5	1\\
-0.5	1.5	1\\
-0.5	2.5	1\\
-0.5	3.5	1\\
-0.5	4.5	1\\
-0.5	5.5	1\\
-0.5	6.5	1\\
-0.5	7.5	1\\
-0.5	8.5	1\\
-0.5	9.5	1\\
-0.5	10.5	1\\
0.5	-0.5	2\\
0.5	0.5	2\\
0.5	1.5	2\\
0.5	2.5	1\\
0.5	3.5	1\\
0.5	4.5	1\\
0.5	5.5	1\\
0.5	6.5	1\\
0.5	7.5	1\\
0.5	8.5	1\\
0.5	9.5	1\\
0.5	10.5	1\\
1.5	-0.5	2\\
1.5	0.5	2\\
1.5	1.5	2\\
1.5	2.5	2\\
1.5	3.5	2\\
1.5	4.5	1\\
1.5	5.5	1\\
1.5	6.5	1\\
1.5	7.5	1\\
1.5	8.5	1\\
1.5	9.5	1\\
1.5	10.5	1\\
2.5	-0.5	2\\
2.5	0.5	2\\
2.5	1.5	2\\
2.5	2.5	2\\
2.5	3.5	2\\
2.5	4.5	2\\
2.5	5.5	2\\
2.5	6.5	1\\
2.5	7.5	1\\
2.5	8.5	1\\
2.5	9.5	1\\
2.5	10.5	1\\
3.5	-0.5	2\\
3.5	0.5	2\\
3.5	1.5	2\\
3.5	2.5	2\\
3.5	3.5	2\\
3.5	4.5	2\\
3.5	5.5	2\\
3.5	6.5	2\\
3.5	7.5	2\\
3.5	8.5	2\\
3.5	9.5	1\\
3.5	10.5	1\\
4.5	-0.5	2\\
4.5	0.5	2\\
4.5	1.5	2\\
4.5	2.5	2\\
4.5	3.5	2\\
4.5	4.5	2\\
4.5	5.5	2\\
4.5	6.5	2\\
4.5	7.5	2\\
4.5	8.5	2\\
4.5	9.5	2\\
4.5	10.5	2\\
5.5	-0.5	2\\
5.5	0.5	2\\
5.5	1.5	2\\
5.5	2.5	2\\
5.5	3.5	2\\
5.5	4.5	2\\
5.5	5.5	2\\
5.5	6.5	2\\
5.5	7.5	2\\
5.5	8.5	2\\
5.5	9.5	2\\
5.5	10.5	2\\
6.5	-0.5	2\\
6.5	0.5	2\\
6.5	1.5	2\\
6.5	2.5	2\\
6.5	3.5	2\\
6.5	4.5	2\\
6.5	5.5	2\\
6.5	6.5	2\\
6.5	7.5	2\\
6.5	8.5	2\\
6.5	9.5	2\\
6.5	10.5	2\\
7.5	-0.5	2\\
7.5	0.5	2\\
7.5	1.5	2\\
7.5	2.5	2\\
7.5	3.5	2\\
7.5	4.5	2\\
7.5	5.5	2\\
7.5	6.5	2\\
7.5	7.5	2\\
7.5	8.5	2\\
7.5	9.5	2\\
7.5	10.5	2\\
8.5	-0.5	2\\
8.5	0.5	2\\
8.5	1.5	2\\
8.5	2.5	2\\
8.5	3.5	2\\
8.5	4.5	2\\
8.5	5.5	2\\
8.5	6.5	2\\
8.5	7.5	2\\
8.5	8.5	2\\
8.5	9.5	2\\
8.5	10.5	2\\
9.5	-0.5	2\\
9.5	0.5	2\\
9.5	1.5	2\\
9.5	2.5	2\\
9.5	3.5	2\\
9.5	4.5	2\\
9.5	5.5	2\\
9.5	6.5	2\\
9.5	7.5	2\\
9.5	8.5	2\\
9.5	9.5	2\\
9.5	10.5	2\\
10.5	-0.5	2\\
10.5	0.5	2\\
10.5	1.5	2\\
10.5	2.5	2\\
10.5	3.5	2\\
10.5	4.5	2\\
10.5	5.5	2\\
10.5	6.5	2\\
10.5	7.5	2\\
10.5	8.5	2\\
10.5	9.5	2\\
10.5	10.5	2\\
};
\end{axis}
\end{tikzpicture}%
}\\
\subfloat[%
  Maximum, $\mathbf{o}\!=\!(0,0)$.
  \label{fig:max00}%
]{
%
%
\begin{tikzpicture}

\begin{axis}[%
width=\mapside,
height=\mapside,
xmin=-0.5,
xmax=10.5,
xlabel={$\Delta_1$},
ylabel={$\Delta_2$},
ymin=-0.5,
ymax=10.5,
xtick
axis background/.style={fill=white},
axis x line*=bottom,
axis y line*=left,
xmajorgrids,
ymajorgrids
]
\addplot[%
surf,
shader=flat corner, draw=black, colormap={mymap}{[1pt] rgb(0pt)=(1,0.694,0.306); rgb(255pt)=(0,0,1)}, mesh/rows=12]
table[row sep=crcr, point meta=\thisrow{c}] {%
x	y	c\\
-0.5	-0.5	2\\
-0.5	0.5	1\\
-0.5	1.5	1\\
-0.5	2.5	1\\
-0.5	3.5	1\\
-0.5	4.5	1\\
-0.5	5.5	1\\
-0.5	6.5	1\\
-0.5	7.5	1\\
-0.5	8.5	1\\
-0.5	9.5	1\\
-0.5	10.5	1\\
0.5	-0.5	2\\
0.5	0.5	2\\
0.5	1.5	2\\
0.5	2.5	1\\
0.5	3.5	1\\
0.5	4.5	1\\
0.5	5.5	1\\
0.5	6.5	1\\
0.5	7.5	1\\
0.5	8.5	1\\
0.5	9.5	1\\
0.5	10.5	1\\
1.5	-0.5	2\\
1.5	0.5	2\\
1.5	1.5	2\\
1.5	2.5	2\\
1.5	3.5	2\\
1.5	4.5	1\\
1.5	5.5	1\\
1.5	6.5	1\\
1.5	7.5	1\\
1.5	8.5	1\\
1.5	9.5	1\\
1.5	10.5	1\\
2.5	-0.5	2\\
2.5	0.5	2\\
2.5	1.5	2\\
2.5	2.5	2\\
2.5	3.5	2\\
2.5	4.5	2\\
2.5	5.5	2\\
2.5	6.5	1\\
2.5	7.5	1\\
2.5	8.5	1\\
2.5	9.5	1\\
2.5	10.5	1\\
3.5	-0.5	2\\
3.5	0.5	2\\
3.5	1.5	2\\
3.5	2.5	2\\
3.5	3.5	2\\
3.5	4.5	2\\
3.5	5.5	2\\
3.5	6.5	2\\
3.5	7.5	2\\
3.5	8.5	2\\
3.5	9.5	1\\
3.5	10.5	1\\
4.5	-0.5	2\\
4.5	0.5	2\\
4.5	1.5	2\\
4.5	2.5	2\\
4.5	3.5	2\\
4.5	4.5	2\\
4.5	5.5	2\\
4.5	6.5	2\\
4.5	7.5	2\\
4.5	8.5	2\\
4.5	9.5	2\\
4.5	10.5	2\\
5.5	-0.5	2\\
5.5	0.5	2\\
5.5	1.5	2\\
5.5	2.5	2\\
5.5	3.5	2\\
5.5	4.5	2\\
5.5	5.5	2\\
5.5	6.5	2\\
5.5	7.5	2\\
5.5	8.5	2\\
5.5	9.5	2\\
5.5	10.5	2\\
6.5	-0.5	2\\
6.5	0.5	2\\
6.5	1.5	2\\
6.5	2.5	2\\
6.5	3.5	2\\
6.5	4.5	2\\
6.5	5.5	2\\
6.5	6.5	2\\
6.5	7.5	2\\
6.5	8.5	2\\
6.5	9.5	2\\
6.5	10.5	2\\
7.5	-0.5	2\\
7.5	0.5	2\\
7.5	1.5	2\\
7.5	2.5	2\\
7.5	3.5	2\\
7.5	4.5	2\\
7.5	5.5	2\\
7.5	6.5	2\\
7.5	7.5	2\\
7.5	8.5	2\\
7.5	9.5	2\\
7.5	10.5	2\\
8.5	-0.5	2\\
8.5	0.5	2\\
8.5	1.5	2\\
8.5	2.5	2\\
8.5	3.5	2\\
8.5	4.5	2\\
8.5	5.5	2\\
8.5	6.5	2\\
8.5	7.5	2\\
8.5	8.5	2\\
8.5	9.5	2\\
8.5	10.5	2\\
9.5	-0.5	2\\
9.5	0.5	2\\
9.5	1.5	2\\
9.5	2.5	2\\
9.5	3.5	2\\
9.5	4.5	2\\
9.5	5.5	2\\
9.5	6.5	2\\
9.5	7.5	2\\
9.5	8.5	2\\
9.5	9.5	2\\
9.5	10.5	2\\
10.5	-0.5	2\\
10.5	0.5	2\\
10.5	1.5	2\\
10.5	2.5	2\\
10.5	3.5	2\\
10.5	4.5	2\\
10.5	5.5	2\\
10.5	6.5	2\\
10.5	7.5	2\\
10.5	8.5	2\\
10.5	9.5	2\\
10.5	10.5	2\\
};
\end{axis}
\end{tikzpicture}%
}
\subfloat[%
  Maximum, $\mathbf{o}\!=\!(0,1)$.
  \label{fig:max01}%
]{
%
%
\begin{tikzpicture}

\begin{axis}[%
width=\mapside,
height=\mapside,
xmin=-0.5,
xmax=10.5,
xlabel={$\Delta_1$},
ylabel={$\Delta_2$},
ymin=-0.5,
ymax=10.5,
xtick
axis background/.style={fill=white},
axis x line*=bottom,
axis y line*=left,
xmajorgrids,
ymajorgrids
]
\addplot[%
surf,
shader=flat corner, draw=black, colormap={mymap}{[255pt] rgb(0pt)=(0,0,1); rgb(255pt)=(1,0.84314,0)}, mesh/rows=12]
table[row sep=crcr, point meta=\thisrow{c}] {%
x	y	c\\
-0.5	-0.5	2\\
-0.5	0.5	2\\
-0.5	1.5	2\\
-0.5	2.5	2\\
-0.5	3.5	2\\
-0.5	4.5	2\\
-0.5	5.5	2\\
-0.5	6.5	2\\
-0.5	7.5	2\\
-0.5	8.5	2\\
-0.5	9.5	2\\
-0.5	10.5	2\\
0.5	-0.5	2\\
0.5	0.5	2\\
0.5	1.5	2\\
0.5	2.5	2\\
0.5	3.5	2\\
0.5	4.5	2\\
0.5	5.5	2\\
0.5	6.5	2\\
0.5	7.5	2\\
0.5	8.5	2\\
0.5	9.5	2\\
0.5	10.5	2\\
1.5	-0.5	2\\
1.5	0.5	2\\
1.5	1.5	2\\
1.5	2.5	2\\
1.5	3.5	2\\
1.5	4.5	2\\
1.5	5.5	2\\
1.5	6.5	2\\
1.5	7.5	2\\
1.5	8.5	2\\
1.5	9.5	2\\
1.5	10.5	2\\
2.5	-0.5	2\\
2.5	0.5	2\\
2.5	1.5	2\\
2.5	2.5	2\\
2.5	3.5	2\\
2.5	4.5	2\\
2.5	5.5	2\\
2.5	6.5	2\\
2.5	7.5	2\\
2.5	8.5	2\\
2.5	9.5	2\\
2.5	10.5	2\\
3.5	-0.5	2\\
3.5	0.5	2\\
3.5	1.5	2\\
3.5	2.5	2\\
3.5	3.5	2\\
3.5	4.5	2\\
3.5	5.5	2\\
3.5	6.5	2\\
3.5	7.5	2\\
3.5	8.5	2\\
3.5	9.5	2\\
3.5	10.5	2\\
4.5	-0.5	2\\
4.5	0.5	2\\
4.5	1.5	2\\
4.5	2.5	2\\
4.5	3.5	2\\
4.5	4.5	2\\
4.5	5.5	2\\
4.5	6.5	2\\
4.5	7.5	2\\
4.5	8.5	2\\
4.5	9.5	2\\
4.5	10.5	2\\
5.5	-0.5	2\\
5.5	0.5	2\\
5.5	1.5	2\\
5.5	2.5	2\\
5.5	3.5	2\\
5.5	4.5	2\\
5.5	5.5	2\\
5.5	6.5	2\\
5.5	7.5	2\\
5.5	8.5	2\\
5.5	9.5	2\\
5.5	10.5	2\\
6.5	-0.5	2\\
6.5	0.5	2\\
6.5	1.5	2\\
6.5	2.5	2\\
6.5	3.5	2\\
6.5	4.5	2\\
6.5	5.5	2\\
6.5	6.5	2\\
6.5	7.5	2\\
6.5	8.5	2\\
6.5	9.5	2\\
6.5	10.5	2\\
7.5	-0.5	2\\
7.5	0.5	2\\
7.5	1.5	2\\
7.5	2.5	2\\
7.5	3.5	2\\
7.5	4.5	2\\
7.5	5.5	2\\
7.5	6.5	2\\
7.5	7.5	2\\
7.5	8.5	2\\
7.5	9.5	2\\
7.5	10.5	2\\
8.5	-0.5	2\\
8.5	0.5	2\\
8.5	1.5	2\\
8.5	2.5	2\\
8.5	3.5	2\\
8.5	4.5	2\\
8.5	5.5	2\\
8.5	6.5	2\\
8.5	7.5	2\\
8.5	8.5	2\\
8.5	9.5	2\\
8.5	10.5	2\\
9.5	-0.5	2\\
9.5	0.5	2\\
9.5	1.5	2\\
9.5	2.5	2\\
9.5	3.5	2\\
9.5	4.5	2\\
9.5	5.5	2\\
9.5	6.5	2\\
9.5	7.5	2\\
9.5	8.5	2\\
9.5	9.5	2\\
9.5	10.5	2\\
10.5	-0.5	2\\
10.5	0.5	2\\
10.5	1.5	2\\
10.5	2.5	2\\
10.5	3.5	2\\
10.5	4.5	2\\
10.5	5.5	2\\
10.5	6.5	2\\
10.5	7.5	2\\
10.5	8.5	2\\
10.5	9.5	2\\
10.5	10.5	2\\
};
\end{axis}
\end{tikzpicture}%
}
\subfloat[%
  Maximum, $\mathbf{o}\!=\!(1,0)$.
  \label{fig:max10}%
]{
%
%
\begin{tikzpicture}

\begin{axis}[%
width=\mapside,
height=\mapside,
xmin=-0.5,
xmax=10.5,
xlabel={$\Delta_1$},
ylabel={$\Delta_2$},
ymin=-0.5,
ymax=10.5,
xtick
axis background/.style={fill=white},
axis x line*=bottom,
axis y line*=left,
xmajorgrids,
ymajorgrids
]
\addplot[%
surf,
shader=flat corner, draw=black, colormap={mymap}{[1pt] rgb(0pt)=(1,0.694,0.306); rgb(255pt)=(0,0,1)}, mesh/rows=12]
table[row sep=crcr, point meta=\thisrow{c}] {%
x	y	c\\
-0.5	-0.5	1\\
-0.5	0.5	1\\
-0.5	1.5	1\\
-0.5	2.5	1\\
-0.5	3.5	1\\
-0.5	4.5	1\\
-0.5	5.5	1\\
-0.5	6.5	1\\
-0.5	7.5	1\\
-0.5	8.5	1\\
-0.5	9.5	1\\
-0.5	10.5	1\\
0.5	-0.5	1\\
0.5	0.5	1\\
0.5	1.5	1\\
0.5	2.5	1\\
0.5	3.5	1\\
0.5	4.5	1\\
0.5	5.5	1\\
0.5	6.5	1\\
0.5	7.5	1\\
0.5	8.5	1\\
0.5	9.5	1\\
0.5	10.5	1\\
1.5	-0.5	1\\
1.5	0.5	1\\
1.5	1.5	1\\
1.5	2.5	1\\
1.5	3.5	1\\
1.5	4.5	1\\
1.5	5.5	1\\
1.5	6.5	1\\
1.5	7.5	1\\
1.5	8.5	1\\
1.5	9.5	1\\
1.5	10.5	1\\
2.5	-0.5	1\\
2.5	0.5	1\\
2.5	1.5	1\\
2.5	2.5	1\\
2.5	3.5	1\\
2.5	4.5	1\\
2.5	5.5	1\\
2.5	6.5	1\\
2.5	7.5	1\\
2.5	8.5	1\\
2.5	9.5	1\\
2.5	10.5	1\\
3.5	-0.5	1\\
3.5	0.5	1\\
3.5	1.5	1\\
3.5	2.5	1\\
3.5	3.5	1\\
3.5	4.5	1\\
3.5	5.5	1\\
3.5	6.5	1\\
3.5	7.5	1\\
3.5	8.5	1\\
3.5	9.5	1\\
3.5	10.5	1\\
4.5	-0.5	1\\
4.5	0.5	1\\
4.5	1.5	1\\
4.5	2.5	1\\
4.5	3.5	1\\
4.5	4.5	1\\
4.5	5.5	1\\
4.5	6.5	1\\
4.5	7.5	1\\
4.5	8.5	1\\
4.5	9.5	1\\
4.5	10.5	1\\
5.5	-0.5	1\\
5.5	0.5	1\\
5.5	1.5	1\\
5.5	2.5	1\\
5.5	3.5	1\\
5.5	4.5	1\\
5.5	5.5	1\\
5.5	6.5	1\\
5.5	7.5	1\\
5.5	8.5	1\\
5.5	9.5	1\\
5.5	10.5	1\\
6.5	-0.5	1\\
6.5	0.5	1\\
6.5	1.5	1\\
6.5	2.5	1\\
6.5	3.5	1\\
6.5	4.5	1\\
6.5	5.5	1\\
6.5	6.5	1\\
6.5	7.5	1\\
6.5	8.5	1\\
6.5	9.5	1\\
6.5	10.5	1\\
7.5	-0.5	1\\
7.5	0.5	1\\
7.5	1.5	1\\
7.5	2.5	1\\
7.5	3.5	1\\
7.5	4.5	1\\
7.5	5.5	1\\
7.5	6.5	1\\
7.5	7.5	1\\
7.5	8.5	1\\
7.5	9.5	1\\
7.5	10.5	1\\
8.5	-0.5	1\\
8.5	0.5	1\\
8.5	1.5	1\\
8.5	2.5	1\\
8.5	3.5	1\\
8.5	4.5	1\\
8.5	5.5	1\\
8.5	6.5	1\\
8.5	7.5	1\\
8.5	8.5	1\\
8.5	9.5	1\\
8.5	10.5	1\\
9.5	-0.5	1\\
9.5	0.5	1\\
9.5	1.5	1\\
9.5	2.5	1\\
9.5	3.5	1\\
9.5	4.5	1\\
9.5	5.5	1\\
9.5	6.5	1\\
9.5	7.5	1\\
9.5	8.5	1\\
9.5	9.5	1\\
9.5	10.5	1\\
10.5	-0.5	1\\
10.5	0.5	1\\
10.5	1.5	1\\
10.5	2.5	1\\
10.5	3.5	1\\
10.5	4.5	1\\
10.5	5.5	1\\
10.5	6.5	1\\
10.5	7.5	1\\
10.5	8.5	1\\
10.5	9.5	1\\
10.5	10.5	1\\
};
\end{axis}
\end{tikzpicture}%
}
\subfloat[%
  Maximum, $\mathbf{o}\!=\!(1,1)$.
  \label{fig:max11}%
]{
%
%
\begin{tikzpicture}

\begin{axis}[%
width=\mapside,
height=\mapside,
xmin=-0.5,
xmax=10.5,
xlabel={$\Delta_1$},
ylabel={$\Delta_2$},
ymin=-0.5,
ymax=10.5,
xtick
axis background/.style={fill=white},
axis x line*=bottom,
axis y line*=left,
xmajorgrids,
ymajorgrids
]
\addplot[%
surf,
shader=flat corner, draw=black, colormap={mymap}{[1pt] rgb(0pt)=(1,0.694,0.306); rgb(255pt)=(0,0,1)}, mesh/rows=12]
table[row sep=crcr, point meta=\thisrow{c}] {%
x	y	c\\
-0.5	-0.5	2\\
-0.5	0.5	2\\
-0.5	1.5	2\\
-0.5	2.5	2\\
-0.5	3.5	2\\
-0.5	4.5	2\\
-0.5	5.5	2\\
-0.5	6.5	2\\
-0.5	7.5	2\\
-0.5	8.5	2\\
-0.5	9.5	2\\
-0.5	10.5	2\\
0.5	-0.5	1\\
0.5	0.5	1\\
0.5	1.5	1\\
0.5	2.5	2\\
0.5	3.5	2\\
0.5	4.5	2\\
0.5	5.5	2\\
0.5	6.5	2\\
0.5	7.5	2\\
0.5	8.5	2\\
0.5	9.5	2\\
0.5	10.5	2\\
1.5	-0.5	1\\
1.5	0.5	1\\
1.5	1.5	1\\
1.5	2.5	1\\
1.5	3.5	1\\
1.5	4.5	2\\
1.5	5.5	2\\
1.5	6.5	2\\
1.5	7.5	2\\
1.5	8.5	2\\
1.5	9.5	2\\
1.5	10.5	2\\
2.5	-0.5	1\\
2.5	0.5	1\\
2.5	1.5	1\\
2.5	2.5	1\\
2.5	3.5	1\\
2.5	4.5	1\\
2.5	5.5	1\\
2.5	6.5	2\\
2.5	7.5	2\\
2.5	8.5	2\\
2.5	9.5	2\\
2.5	10.5	2\\
3.5	-0.5	1\\
3.5	0.5	1\\
3.5	1.5	1\\
3.5	2.5	1\\
3.5	3.5	1\\
3.5	4.5	1\\
3.5	5.5	1\\
3.5	6.5	1\\
3.5	7.5	1\\
3.5	8.5	1\\
3.5	9.5	2\\
3.5	10.5	2\\
4.5	-0.5	1\\
4.5	0.5	1\\
4.5	1.5	1\\
4.5	2.5	1\\
4.5	3.5	1\\
4.5	4.5	1\\
4.5	5.5	1\\
4.5	6.5	1\\
4.5	7.5	1\\
4.5	8.5	1\\
4.5	9.5	1\\
4.5	10.5	1\\
5.5	-0.5	1\\
5.5	0.5	1\\
5.5	1.5	1\\
5.5	2.5	1\\
5.5	3.5	1\\
5.5	4.5	1\\
5.5	5.5	1\\
5.5	6.5	1\\
5.5	7.5	1\\
5.5	8.5	1\\
5.5	9.5	1\\
5.5	10.5	1\\
6.5	-0.5	1\\
6.5	0.5	1\\
6.5	1.5	1\\
6.5	2.5	1\\
6.5	3.5	1\\
6.5	4.5	1\\
6.5	5.5	1\\
6.5	6.5	1\\
6.5	7.5	1\\
6.5	8.5	1\\
6.5	9.5	1\\
6.5	10.5	1\\
7.5	-0.5	1\\
7.5	0.5	1\\
7.5	1.5	1\\
7.5	2.5	1\\
7.5	3.5	1\\
7.5	4.5	1\\
7.5	5.5	1\\
7.5	6.5	1\\
7.5	7.5	1\\
7.5	8.5	1\\
7.5	9.5	1\\
7.5	10.5	1\\
8.5	-0.5	1\\
8.5	0.5	1\\
8.5	1.5	1\\
8.5	2.5	1\\
8.5	3.5	1\\
8.5	4.5	1\\
8.5	5.5	1\\
8.5	6.5	1\\
8.5	7.5	1\\
8.5	8.5	1\\
8.5	9.5	1\\
8.5	10.5	1\\
9.5	-0.5	1\\
9.5	0.5	1\\
9.5	1.5	1\\
9.5	2.5	1\\
9.5	3.5	1\\
9.5	4.5	1\\
9.5	5.5	1\\
9.5	6.5	1\\
9.5	7.5	1\\
9.5	8.5	1\\
9.5	9.5	1\\
9.5	10.5	1\\
10.5	-0.5	1\\
10.5	0.5	1\\
10.5	1.5	1\\
10.5	2.5	1\\
10.5	3.5	1\\
10.5	4.5	1\\
10.5	5.5	1\\
10.5	6.5	1\\
10.5	7.5	1\\
10.5	8.5	1\\
10.5	9.5	1\\
10.5	10.5	1\\
};
\end{axis}
\end{tikzpicture}%
}
    \caption{Optimal policies for the count range query (upper) and maximum query (lower) with $p_1=0.1$ and $p_2=0.2$.}
    \label{fig:policies}
\end{figure*}

The results for $p_1=0.1$ and $p_2=0.2$ are given in Fig.~\ref{fig:policies}. As Fig.~\ref{fig:cnt00}-\subref*{fig:cnt11} show, the policy is the same for any observation, and only depends on the age of the two measurements, since the \gls{mse} of the count query is the same for any observation. The level of uncertainty determines the action: the second component of the state, which can vary more often due to the higher state change probability, is the one that is polled, unless the age of the latest observation of the other component is approximately double. The maximum query has a more complex policy: if the last observations of each component are the same, i.e., $\mathbf{o}=(0,0)$ or $\mathbf{o}=(1,1)$, the policy is the same as for the count range query. On the other hand, if one of the last observations is 1, while the other is 0, the component with value 1 is always polled. This is reasonable, as one observation from a sensor that contains a 1 gives perfect certainty on the overall maximum. Giving a higher priority to the component with the highest probability of being equal to 1 is then beneficial, even if the uncertainty on the other component becomes extremely high.

Naturally, this is only a simple example, and introducing a query process will complicate the system, but it highlights the strong dependence between the function that determines each query and the respective polling policy. While the optimal strategy to minimize the \gls{aoi} would always poll the sensor with the highest age, and a strategy that minimizes the uncertainty of the count range query (which, in this simple system, is almost equivalent to minimizing the \gls{mse}) weighs each sensor's age by the speed of the corresponding process, the strategy for the maximum query actually depends on the current value of each sensor, and is starkly  different from the others. Mixing different types of query in the same system will then lead to non-trivial trade-offs, particularly when the functions are highly non-linear.

\subsection{Reinforcement Learning Solution and Learning Architecture}\label{ssec:rl}
While policy iteration has strong convergence guarantees, it is infeasible to use when the state space is large, which is the case for the considered scheduling problem. Instead, we resort to approximate solutions, and consider a \gls{rl} approach to the scheduling problem. \gls{rl} is a machine learning approach in which an agent learns from experience, updating its estimate of the value function by trial and error. The agent makes decisions and receives immediate rewards from the environment, without any prior knowledge of the reward function or the consequences of actions. For a more thorough introduction to reinforcement learning, we refer the reader to \cite{sutton2018reinforcement}. While \gls{rl} is not directly applicable to partially observable scenarios, the observation space in the full problem is a sufficient statistic to respect the Markov property: as the Kalman filter is the optimal estimator for linear systems, and all the historic information of past observations is available in the state and covariance estimates, the system is equivalent to a fully observable \gls{mdp} and we can directly apply standard solutions such as \gls{rl}~\cite{hansen1997improved}.

We implement the \gls{dqn} architecture~\cite{mnih2015human}, which uses a deep neural network to approximate the value function. In order to avoid instability, we need to use a \emph{replay memory} to store the agent experience and select batches of uncorrelated samples. Each batch contains $B$ uncorrelated samples, and each experience sample is a tuple $e=(s(t),a(t),r(t),s(t+1))$. We maintain two neural networks for increased stability: a \emph{target network} and an \emph{update network}. In order to estimate the long-term reward $R(\pi)$ from an experience sample, we use the target network's prediction $Q_t(s,a)$:
\begin{equation}
 Q(e)=r(t)+\gamma \max_{a\in\mathcal{A}}Q_t(s(t+1),a).
\end{equation}
The use of the long-term reward estimates to update future estimates follows the well-established bootstrap method, and the use of a greedy update policy follows the Q-learning model implemented by the \gls{dqn}. The estimates $Q(e)$ are then used as labels for the backpropagation operation on the update network, whose output predictions are used in the action policy to select the next action. The action policy we use implements the well-known softmax function:
\begin{equation}
 \pi(s,a)=\frac{e^{\frac{Q_u(s,a)}{\tau}}}{\sum_{a'\in\mathcal{A}}e^{\frac{Q_u(s,a')}{\tau}}},
\end{equation}
where the temperature parameter $\tau$ is to balance between exploration and exploitation. Lower values of $\tau$ make the outcome closer to the greedy policy, as the probability of selecting suboptimal actions decreases, while higher values of $\tau$ increase exploration. In any case, exploration with the softmax function is \emph{directed}: actions that are assumed to be highly suboptimal will be picked less frequently, while the agent will prefer actions that have an estimated long-term reward just below the maximum.

The update network is updated at every step with a new batch of samples, while the target network is only updated every $U$ steps by copying the update network's weights. As we stated above, the use of separate target and update networks allows the system to converge, avoiding numerical and stability issues. In the rest of this work, we implement a \gls{dqn} with 3 layers, whose parameters are given in Table~\ref{tab:learning}. The first two layers have a dropout probability $p_d=0.1$ during the training, and the network is relatively simple, as the input is highly redundant. The hyperparameters above were found after a grid search optimization process.

\begin{table}[t!]
\centering
	\caption{\gls{dqn} architecture.}
	\label{tab:learning}
    \scriptsize
	\begin{tabular}[c]{c|ccc}
		\toprule
		Parameter & Layer 1 & Layer 2 & Layer 3 \\
		\midrule
		Input size & $M^2+M+C$ & $2.5M$ & $M$\\
		Output size & $2.5M$ & $M$ & $N$\\
		Activation function & ReLU & ReLU & ReLU\\
		Dropout & 0.1 & 0.1 & 0\\
		\bottomrule
	\end{tabular}
\end{table}

\subsection{Computational Complexity}

We can now discuss the computational complexity of the learning solution. The following refers to the complexity of a trained model, i.e., of a single decision on the next action: while training can be performed offline in a simulation environment or even passively on existing data, actions need to be real-time for the system to work, and the time to make decisions is critical.

If we consider a single layer with $\ell_i$ inputs and $\ell_o$ outputs, there are three operations that the network needs to perform to compute each output:
\begin{enumerate}
 \item Multiply each input $\ell_i$ by the appropriate weight (equivalent to $\ell_i$ multiplication operations);
 \item Sum all the results (equivalent to $\ell_i$ sums);
 \item Apply the non-linear activation function.
\end{enumerate}
If we consider the activation function as the result of $k$ basic operations, the total number of basic operations for a single layer is then $\ell_o(2\ell_i+k)$. If we consider our architecture as a vector $\bm{\ell}$ of layer sizes, where the first element is the cardinality of the input (i.e., the observed state) and the last element is $N$ (corresponding to the $N$ possible actions), the total complexity is:
\begin{equation}
 \mathcal{C}_f(\bm{\ell})=\sum_{i=1}^{|\bm{\ell}|-1}\ell_{i+1}(2\ell_i+k).
\end{equation}
We remark that $\mathcal{C}_f(\bm{\ell})$ is the total number of basic operations per layer, and as such, 
If we consider our architecture for $C=2$ and $N=20$, which is given above, we have $k=1$, as the \gls{relu} activation function is extremely simple, and the total number of operations for each step is then $\mathcal{C}_f=96\,570$.
The backpropagation algorithm required to train the neural network has the same complexity as the forward pass, but it must be run for each sample in a training batch~\cite{widrow199030}. For a training batch size of $B$, the total complexity for a single training step is then given by:
\begin{equation}
 \mathcal{C}_b(\bm{\ell})=B\mathcal{C}_f(\bm{\ell}).
\end{equation}
In our architecture, we have $B=128$, and consequently, $\mathcal{C}_b=12\,360\,960$. This number of operations should be entirely within the capabilities of an edge node, as even simple embedded processors can deal with much more complex architectures that require billions of operations in less than 100 ms~\cite{bianco2018benchmark}. As most of the required calculations in training and evaluation are vector operations, each layer might only require a single clock tick on modern processors, particularly when the processor is a GPU or designed for hardware-assisted learning.

\section{Simulation Settings and Results}\label{sec:sim}
The performance of the \gls{rl}-based query-aware scheme is verified by Monte Carlo simulation, considering a specific scenario. Its performance is measured in terms of the \gls{mse} on its query responses. The evaluation is performed over $E_{\text{test}}=10$ independent episodes, each of which consists of $T_{\text{max}}=100$ time steps. The parameters of the \gls{dqn} agent are the same for all considered scenarios, and are given in Table~\ref{tab:dqn_param}.

\begin{table}[t!]
\centering
	\caption{\gls{dqn} parameters.}
	\label{tab:dqn_param}
    \scriptsize
	\begin{tabular}[c]{clc}
		\toprule
		Parameter & Description & Value \\
		\midrule
		$\gamma$ & Exponential discount factor & 0.9\\
		$T_e$ & Time steps in each episode & 100\\
		$E_{\text{train}}$ & Training episodes & 100\\
		$E_{\text{test}}$ & Test episodes & 10\\
		$p_d$ & Dropout probability & 0.1\\
		$R_m$ & Replay memory size & 10000\\
		$B$ & Batch size & 128\\
		$t_{\text{up}}$ & Target net update period & 10\\
		$L_o$& Learning rate optimizer & Adam\\
		$L_0$ & Initial learning rate & $10^{-4}$\\
		\bottomrule
	\end{tabular}
\end{table}

\subsection{Scenario and Benchmark Policies}
We consider a system with $M=20$ sensors, each observing a different component of the state $\mathbf{x}(t)$, so that $M=N$ and $\mathbf{H}=\mathbf{I}$.
The dynamic system that the edge node observes is defined as follows:
\begin{equation}
    \mathbf{A}^{(i,j)}=\begin{cases}
            \frac{3}{4}, &\text{if } i=j;\\
            -\frac{1}{8}, &\text{if } i\neq j\wedge \text{mod}(i-2j,7)=6.
          \end{cases}
\end{equation}
The edge node knows $\mathbf{A}$, as well as the process and measurement noise covariance matrices, which are given by:
\begin{align}
\bm{\Sigma}_v^{(i,j)}&=\begin{cases}
             \frac{11+\text{mod}(i-1,10)}{5}, &\text{if } i=j;\\
             1, &\text{if } i\neq j, \text{mod}(i-j,6)=0,
           \end{cases}\\
           \bm{\Sigma}_w&=\mathbf{I}.
\end{align}
The error probability $\varepsilon_n$ for each sensor is given by:
\begin{equation}
    \varepsilon_n=0.02\left\lceil\frac{n}{10}\right\rceil.
\end{equation}
We consider a case with $C=2$ clients with the same importance, i.e., $\alpha_1=\alpha_2=1$. Client 1 requests a \emph{count range} query with interval $[-5,0]$, while client 2 makes a \emph{maximum} query. The two query types are described in more detail in Sec.~\ref{ssec:queries}, and we also refer the reader to our previous work~\cite{chiariotti2022scheduling} for a deeper discussion on the derivation of \gls{mmse} estimators for specific queries.
It is possible to add more clients with other queries and varying importance. This adds one parameter (i.e., the time since the last query from that client) to the \gls{dqn}, but the problem is not guaranteed to scale: the added complexity necessarily makes the training longer, requiring an adjustment to the exploration and learning profiles as well. In this work, we also include the results for a scenario with 4 clients, including an \emph{average} and a \emph{state} query as well.

We consider 5 different benchmarks for the query-aware policy:
\begin{itemize}
    \item \emph{\gls{maf}}: The \gls{maf} policy, which minimizes the average \gls{aoi} of the system regardless of the value of sensors' readings. This legacy approach represents a value-neutral lower bound, as it aims at minimizing the \gls{aoi} for all sensors regardless of the relevance of their data or their expected effect on the accuracy of the state estimate;
    \item \emph{Cnt}: The one-step optimal policy for client 1, which follows the procedure from~\cite{chiariotti2022scheduling} to minimize the \gls{mse} of the count range query in the current step;
    \item \emph{Max}: The one-step optimal policy for client 2, which does the same for the maximum query;
    \item \emph{\gls{rl} (Cnt)}: The foresighted policy learned by a \gls{rl} agent with $\alpha_2=0$, which only minimizes the \gls{mse} of the response to the count range query;
    \item \emph{\gls{rl} (Max)}: The foresighted policy with $\alpha_1=0$, which does the same for the maximum query.
\end{itemize}
We also consider two different query processes, both of which are observable by the edge node, slightly simplifying the problem:
\begin{itemize}
    \item \emph{Periodic queries}: queries are generated every $T_q=6$ steps. In this case, the Markov chain is deterministic, going from state 0 (in which a query is generated) to state 1 with probability 1, then increasing until 5, after which the chain goes back to 0 with probability 1;
    \item \emph{Memoryless queries}: in this case, the Markov chain only has 2 states, and the rows of the transition matrix are identical. The time between two subsequent queries is geometrically distributed, with an expected value $\mathbb{E}[T_q]=6$ steps.
\end{itemize}
We consider three combinations of these query processes: the case in which both clients have periodic queries, the case in which they both have memoryless queries, and the mixed case in which client 1 has periodic queries, while client 2 follows a memoryless process.

\begin{figure}[t]
    \centering
    \input{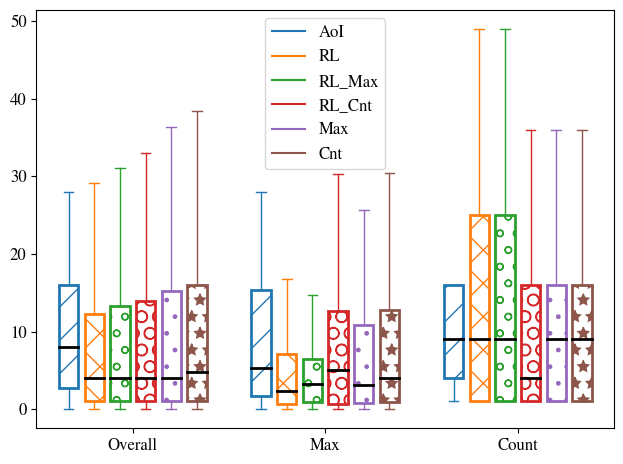}
    \caption{\gls{mse} cost of the different policies for both types of query in the periodic query scenario.}
    \label{fig:Periodic_box}
\end{figure}

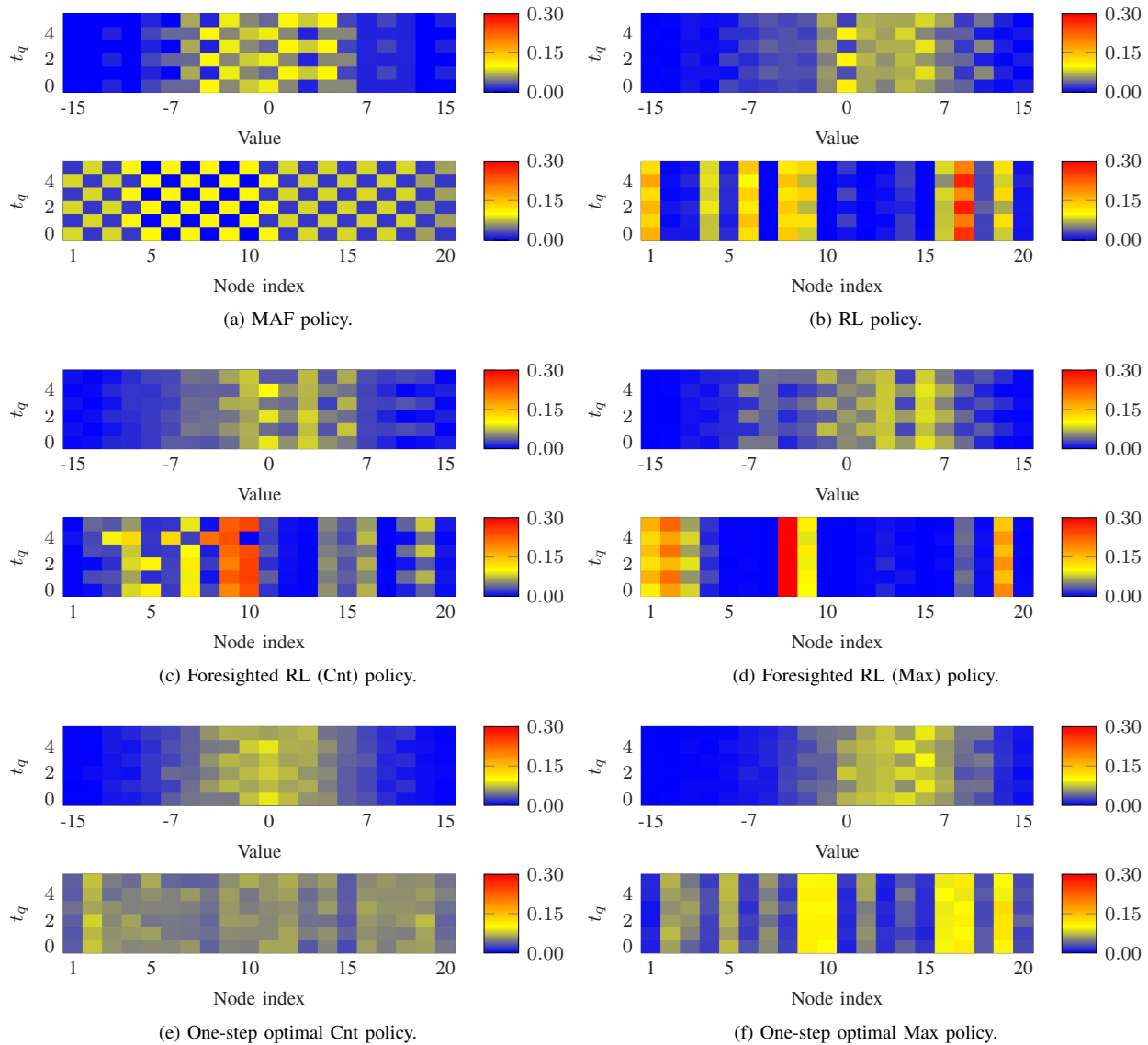
\begin{figure*}[t]
\centering
    \subfloat[%
  MAF policy. \label{fig:color_maf}%
]{\begin{tikzpicture}
  \begin{groupplot}[
    group style={group size=1 by 2,
    vertical sep=1cm,
    x descriptions at=edge bottom,
    y descriptions at=edge left},
  enlargelimits=false, ymax=5.5,
  height=\cmheight,width=\cmwidth]
\nextgroupplot[point meta min=0,
xtick={0,5,10,15,19},
xticklabels={-15,-7,0,7,15},
point meta max=0.3,
ylabel=$t_q$,
xlabel=Value,
    colorbar,
    colorbar style={%
      ymax=0.3,
      ytick={0.0,0.15,0.30},
      /pgf/number format/precision=2,
      /pgf/number format/fixed,
      /pgf/number format/fixed zerofill}]
\addplot [
matrix plot*,
mesh/cols=20,
point meta=explicit,
] table [meta index=2] {Color_plots/AoI_data.csv};
\nextgroupplot[%
  point meta min=0,
  point meta max=0.3,
ylabel=$t_q$,
  xlabel=Node index,
    xtick={0,4,9,14,19},
    xticklabels={1,5,10,15,20},
  colorbar,
  colorbar style={%
  ymax=0.3,
    ytick={0.00,0.15,0.30},
    /pgf/number format/precision=2,
    /pgf/number format/fixed,
    /pgf/number format/fixed zerofill}]
  \addplot [
  matrix plot*,
  mesh/cols=20,
  point meta=explicit,
  ] table [meta index=2] {Color_plots/AoI_actions.csv};
\end{groupplot}
\end{tikzpicture}}
    \subfloat[%
  RL policy. \label{fig:color_rl}%
]{\begin{tikzpicture}
  \begin{groupplot}[
    group style={group size=1 by 2,
    vertical sep=1cm,
    x descriptions at=edge bottom,
    y descriptions at=edge left},
  enlargelimits=false, ymax=5.5,
  height=\cmheight,width=\cmwidth]
\nextgroupplot[point meta min=0,
xtick={0,5,10,15,19},
xticklabels={-15,-7,0,7,15},
point meta max=0.3,
ylabel=$t_q$,
xlabel=Value,
    colorbar,
    colorbar style={%
      ymax=0.3,
      ytick={0.0,0.15,0.30},
      /pgf/number format/precision=2,
      /pgf/number format/fixed,
      /pgf/number format/fixed zerofill}]
\addplot [
matrix plot*,
mesh/cols=20,
point meta=explicit,
] table [meta index=2] {Color_plots/RL_data.csv};
\nextgroupplot[%
  point meta min=0,
  point meta max=0.3,
ylabel=$t_q$,
  xlabel=Node index,
    xtick={0,4,9,14,19},
    xticklabels={1,5,10,15,20},
  colorbar,
  colorbar style={%
  ymax=0.3,
    ytick={0.00,0.15,0.30},
    /pgf/number format/precision=2,
    /pgf/number format/fixed,
    /pgf/number format/fixed zerofill}]
  \addplot [
  matrix plot*,
  mesh/cols=20,
  point meta=explicit,
  ] table [meta index=2] {Color_plots/RL_actions.csv};
\end{groupplot}
\end{tikzpicture}}\\
    \subfloat[%
  Foresighted \gls{rl} (Cnt) policy. \label{fig:color_rlcnt}%
]{\begin{tikzpicture}
  \begin{groupplot}[
    group style={group size=1 by 2,
    vertical sep=1cm,
    x descriptions at=edge bottom,
    y descriptions at=edge left},
  enlargelimits=false, ymax=5.5,
  height=\cmheight,width=\cmwidth]
\nextgroupplot[point meta min=0,
xtick={0,5,10,15,19},
xticklabels={-15,-7,0,7,15},
point meta max=0.3,
ylabel=$t_q$,
xlabel=Value,
    colorbar,
    colorbar style={%
      ymax=0.3,
      ytick={0.0,0.15,0.30},
      /pgf/number format/precision=2,
      /pgf/number format/fixed,
      /pgf/number format/fixed zerofill}]
\addplot [
matrix plot*,
mesh/cols=20,
point meta=explicit,
] table [meta index=2] {Color_plots/RL_Cnt_data.csv};
\nextgroupplot[%
  point meta min=0,
  point meta max=0.3,
ylabel=$t_q$,
  xlabel=Node index,
    xtick={0,4,9,14,19},
    xticklabels={1,5,10,15,20},
  colorbar,
  colorbar style={%
  ymax=0.3,
    ytick={0.00,0.15,0.30},
    /pgf/number format/precision=2,
    /pgf/number format/fixed,
    /pgf/number format/fixed zerofill}]
  \addplot [
  matrix plot*,
  mesh/cols=20,
  point meta=explicit,
  ] table [meta index=2] {Color_plots/RL_Cnt_actions.csv};
\end{groupplot}
\end{tikzpicture}}
    \subfloat[%
  Foresighted \gls{rl} (Max) policy. \label{fig:color_rlmax}%
]{\begin{tikzpicture}
  \begin{groupplot}[
    group style={group size=1 by 2,
    vertical sep=1cm,
    x descriptions at=edge bottom,
    y descriptions at=edge left},
  enlargelimits=false, ymax=5.5,
  height=\cmheight,width=\cmwidth]
\nextgroupplot[point meta min=0,
xtick={0,5,10,15,19},
xticklabels={-15,-7,0,7,15},
point meta max=0.3,
ylabel=$t_q$,
xlabel=Value,
    colorbar,
    colorbar style={%
      ymax=0.3,
      ytick={0.0,0.15,0.30},
      /pgf/number format/precision=2,
      /pgf/number format/fixed,
      /pgf/number format/fixed zerofill}]
\addplot [
matrix plot*,
mesh/cols=20,
point meta=explicit,
] table [meta index=2] {Color_plots/RL_Max_data.csv};
\nextgroupplot[%
  point meta min=0,
  point meta max=0.3,
ylabel=$t_q$,
    xtick={0,4,9,14,19},
    xticklabels={1,5,10,15,20},
  xlabel=Node index,
  colorbar,
  colorbar style={%
  ymax=0.3,
    ytick={0.00,0.15,0.30},
    /pgf/number format/precision=2,
    /pgf/number format/fixed,
    /pgf/number format/fixed zerofill}]
  \addplot [
  matrix plot*,
  mesh/cols=20,
  point meta=explicit,
  ] table [meta index=2] {Color_plots/RL_Max_actions.csv};
\end{groupplot}
\end{tikzpicture}}\\
    \subfloat[%
  One-step optimal Cnt policy. \label{fig:color_cnt}%
]{\begin{tikzpicture}
  \begin{groupplot}[
    group style={group size=1 by 2,
    vertical sep=1cm,
    x descriptions at=edge bottom,
    y descriptions at=edge left},
  enlargelimits=false, ymax=5.5,
  height=\cmheight,width=\cmwidth]
\nextgroupplot[point meta min=0,
xtick={0,5,10,15,19},
xticklabels={-15,-7,0,7,15},
point meta max=0.3,
ylabel=$t_q$,
xlabel=Value,
    colorbar,
    colorbar style={%
      ymax=0.3,
      ytick={0.0,0.15,0.30},
      /pgf/number format/precision=2,
      /pgf/number format/fixed,
      /pgf/number format/fixed zerofill}]
\addplot [
matrix plot*,
mesh/cols=20,
point meta=explicit,
] table [meta index=2] {Color_plots/Cnt_data.csv};
\nextgroupplot[%
  point meta min=0,
  point meta max=0.3,
ylabel=$t_q$,
  xlabel=Node index,
    xtick={0,4,9,14,19},
    xticklabels={1,5,10,15,20},
  colorbar,
  colorbar style={%
  ymax=0.3,
    ytick={0.00,0.15,0.30},
    /pgf/number format/precision=2,
    /pgf/number format/fixed,
    /pgf/number format/fixed zerofill}]
  \addplot [
  matrix plot*,
  mesh/cols=20,
  point meta=explicit,
  ] table [meta index=2] {Color_plots/Cnt_actions.csv};
\end{groupplot}
\end{tikzpicture}}
    \subfloat[%
  One-step optimal Max policy. \label{fig:color_max}%
]{\begin{tikzpicture}
  \begin{groupplot}[
    group style={group size=1 by 2,
    vertical sep=1cm,
    x descriptions at=edge bottom,
    y descriptions at=edge left},
  enlargelimits=false, ymax=5.5,
  height=\cmheight,width=\cmwidth]
\nextgroupplot[point meta min=0,
xtick={0,5,10,15,19},
xticklabels={-15,-7,0,7,15},
point meta max=0.3,
ylabel=$t_q$,
xlabel=Value,
    colorbar,
    colorbar style={%
      ymax=0.3,
      ytick={0.0,0.15,0.30},
      /pgf/number format/precision=2,
      /pgf/number format/fixed,
      /pgf/number format/fixed zerofill}]
\addplot [
matrix plot*,
mesh/cols=20,
point meta=explicit,
] table [meta index=2] {Color_plots/Max_data.csv};
\nextgroupplot[%
  point meta min=0,
  point meta max=0.3,
ylabel=$t_q$,
    xtick={0,4,9,14,19},
    xticklabels={1,5,10,15,20},
  xlabel=Node index,
  colorbar,
  colorbar style={%
  ymax=0.3,
    ytick={0.00,0.15,0.30},
    /pgf/number format/precision=2,
    /pgf/number format/fixed,
    /pgf/number format/fixed zerofill}]
  \addplot [
  matrix plot*,
  mesh/cols=20,
  point meta=explicit,
  ] table [meta index=2] {Color_plots/Max_actions.csv};
\end{groupplot}
\end{tikzpicture}}
    \caption{Colormap representing the histogram of choices, in terms of sensor value and index, for each of the policies. The vertical axis represents the time $t_q$ (count range queries arrive at time 0).}
    \label{fig:colormap_rl}
\end{figure*}

\subsection{Periodic Query Scenario}\label{ssec:periodic}
In this subsection, we show results for the case where both queries arrive periodically, each with a period of 6 steps. The two queries are out of sync, with the maximum query starting at 0 and the count range query starting at time 2.

This is the easiest case for the edge node, as queries are generated deterministically and the optimal policy can act on deterministic knowledge of the query pattern. Fig.~\ref{fig:Periodic_box} shows a boxplot of the \gls{mse} for both types of queries, as well as the overall cost, which is defined as the opposite of the average reward as given in~\eqref{eq:reward}. In other words, the overall cost is a weighted sum of the \glspl{mse} for all queries, using the weight vector $\bm{\alpha}$. In our simulations, queries all have the same period and weight, so the overall cost corresponds to the average of the \glspl{mse} over all queries. We can note that the \gls{rl} policy considering both types of queries obtains a lower cost than all others (as shown in the group on the left of the figure), with a much lower average and only slightly worse performance at the 95th percentile than an \gls{aoi}-oriented approach. In particular, the choice made by the \gls{rl} policy is to privilege the maximum query, with results that end up being similar to only optimizing for it. All other approaches tend to reduce the \gls{mse} of the count range query more, although they end up having a higher error on the maximum query. The count range query is penalized by the fact that it arrives only 2 slots after the maximum query: reducing its \gls{mse} would require losing accuracy in the response to the maximum query, increasing the overall cost. On the other hand, the 4 slots between a count range query and the subsequent maximum query allow the \gls{rl} policy to improve the accuracy significantly. The effect of the discount factor $\gamma$ is also important: since a count range query arrives 2 slots after the maximum query before it, and $\gamma=0.9$, its \gls{mse} only accounts for 81\% of the reward for the steps before the maximum query. A higher value of $\gamma$, or a different weighting of the two query types by adapting $\alpha_1$, would produce a more balanced outcome.

We can also note that, in this case, the other \gls{rl}-based policies outperform their greedy versions on the metric that they optimize for, but no such pattern exists for the other type of query, which these policies completely disregard. As noted in previous works on \gls{voi}, the \gls{aoi}-based approach taken by the \gls{maf} policy provides a middle ground for performance, never failing too badly by polling all sensors equally. The choices made by the various policies can be analyzed more in depth by considering the distribution of the sensors that are polled. Fig.~\ref{fig:colormap_rl} shows two colormaps for each policy, in which the y-axis represents the step in each query period, i.e., the index of the slot modulus 6. As a reminder, the maximum query is generated at $t_q=0$ and the count range query is generated at $t_q=2$. The two colormaps differ by the value represented on the x-axis: in the first one, the x-axis represents the value $x^{(m)}(t)$ measured by the chosen sensor, while in the second, the value is simply the index of the sensor. The color of each cell represents the empirical probability of each combination in our test episodes.

\begin{figure*}[t]
    \centering
    \centering
    \subfloat[%
  Average AoI for each node. \label{fig:aoi_eval}%
]{\begin{tikzpicture}
  \begin{groupplot}[
    group style={group size=1 by 2,
    vertical sep=1cm,
    x descriptions at=edge bottom,
    y descriptions at=edge left},
    scale only axis,
  enlargelimits=false, ymax=5.5,
  height=0.2\linewidth,width=0.35\linewidth]
\nextgroupplot[point meta min=0,
xtick={0,6,9,14,19},
xticklabels={1,5,10,15,20},
tick align=outside,
tick pos=left,
ytick={0,1,2,3,4,5},
yticklabels={MAF,RL,RL (Max),RL (Cnt),Max,Cnt},
point meta max=50,
xlabel=Node index,
    colorbar,
    colorbar style={%
      ymax=50,
      ytick={0,10,20,30,40,50}}]
\addplot [
matrix plot*,
mesh/cols=20,
point meta=explicit,
] table [meta index=2] {Color_plots/Average_Age.csv};
\end{groupplot}
\end{tikzpicture}}
    \subfloat[%
  State and query MSE for all policies. \label{fig:mse_eval}%
]{
\begin{tikzpicture}

\begin{axis}[
width=0.35\linewidth,
height=0.2\linewidth,
scale only axis,
legend cell align={left},
legend style={
  fill opacity=0.8,
  draw opacity=1,
  text opacity=1,
  at={(0.5,0.98)},
  anchor=north,
  draw=white!80!black
},
tick align=outside,
tick pos=left,
ylabel={MSE},
x grid style={white!69.0196078431373!black},
xmin=-3, xmax=9,
xtick style={color=black},
xtick={0,6},
xlabel={\textcolor{white}{a}},
xticklabels={State MSE, Overall cost},
y grid style={white!69.0196078431373!black},
ymin=0, ymax=50,
ytick style={color=black},
legend columns=3,
legend entries={MAF,RL,RL (Max),RL (Cnt),Max,Cnt}
]
\addplot [color0, forget plot]
table {%
-2 6.42375211725632
-2 0
};
\addplot [color0, forget plot]
table {%
-2 11.9768267771577
-2 14.9955056898416
};
\addplot [color0, forget plot]
table {%
-2.15 0
-1.85 0
};
\addplot [color0, forget plot]
table {%
-2.15 14.9955056898416
-1.85 14.9955056898416
};
\addplot [color0, forget plot]
table {%
4 2.79273544534273
4 0.0190467944036528
};
\addplot [color0, forget plot]
table {%
4 16
4 27.9669135379411
};
\addplot [color0, forget plot]
table {%
3.85 0.0190467944036528
4.15 0.0190467944036528
};
\addplot [color0, forget plot]
table {%
3.85 27.9669135379411
4.15 27.9669135379411
};
\addplot [color1, forget plot]
table {%
-1.2 8.00168372464032
-1.2 0
};
\addplot [color1, forget plot]
table {%
-1.2 14.2655927728142
-1.2 23.6563111777846
};
\addplot [color1, forget plot]
table {%
-1.35 0
-1.05 0
};
\addplot [color1, forget plot]
table {%
-1.35 23.6563111777846
-1.05 23.6563111777846
};
\addplot [color1, forget plot]
table {%
4.8 1
4.8 5.93690251963829e-07
};
\addplot [color1, forget plot]
table {%
4.8 12.4497512684613
4.8 29.6096217790708
};
\addplot [color1, forget plot]
table {%
4.65 5.93690251963829e-07
4.95 5.93690251963829e-07
};
\addplot [color1, forget plot]
table {%
4.65 29.6096217790708
4.95 29.6096217790708
};
\addplot [color2, forget plot]
table {%
-0.4 8.15664958362689
-0.4 0
};
\addplot [color2, forget plot]
table {%
-0.4 15.4991117226915
-0.4 26.4823834339294
};
\addplot [color2, forget plot]
table {%
-0.55 0
-0.25 0
};
\addplot [color2, forget plot]
table {%
-0.55 26.4823834339294
-0.25 26.4823834339294
};
\addplot [color2, forget plot]
table {%
5.6 1
5.6 1.41993579294643e-05
};
\addplot [color2, forget plot]
table {%
5.6 13.4981681993061
5.6 30.3421472818926
};
\addplot [color2, forget plot]
table {%
5.45 1.41993579294643e-05
5.75 1.41993579294643e-05
};
\addplot [color2, forget plot]
table {%
5.45 30.3421472818926
5.75 30.3421472818926
};
\addplot [color3, forget plot]
table {%
0.4 7.91783402445769
0.4 0
};
\addplot [color3, forget plot]
table {%
0.4 14.1303506990947
0.4 23.4381671699959
};
\addplot [color3, forget plot]
table {%
0.25 0
0.55 0
};
\addplot [color3, forget plot]
table {%
0.25 23.4381671699959
0.55 23.4381671699959
};
\addplot [color3, forget plot]
table {%
6.4 1
6.4 1.02927009531706e-05
};
\addplot [color3, forget plot]
table {%
6.4 13.8835360771815
6.4 33.206613828342
};
\addplot [color3, forget plot]
table {%
6.25 1.02927009531706e-05
6.55 1.02927009531706e-05
};
\addplot [color3, forget plot]
table {%
6.25 33.206613828342
6.55 33.206613828342
};
\addplot [color4, forget plot]
table {%
1.2 6.850641937635
1.2 0
};
\addplot [color4, forget plot]
table {%
1.2 12.4711901401648
1.2 20.8919567865094
};
\addplot [color4, forget plot]
table {%
1.05 0
1.35 0
};
\addplot [color4, forget plot]
table {%
1.05 20.8919567865094
1.35 20.8919567865094
};
\addplot [color4, forget plot]
table {%
7.2 1
7.2 2.29723606988603e-07
};
\addplot [color4, forget plot]
table {%
7.2 15.2751198980687
7.2 36.3419106482347
};
\addplot [color4, forget plot]
table {%
7.05 2.29723606988603e-07
7.35 2.29723606988603e-07
};
\addplot [color4, forget plot]
table {%
7.05 36.3419106482347
7.35 36.3419106482347
};
\addplot [color5, forget plot]
table {%
2 6.60707000627882
2 0
};
\addplot [color5, forget plot]
table {%
2 11.4620940090214
2 18.7395139903031
};
\addplot [color5, forget plot]
table {%
1.85 0
2.15 0
};
\addplot [color5, forget plot]
table {%
1.85 18.7395139903031
2.15 18.7395139903031
};
\addplot [color5, forget plot]
table {%
8 1
8 1.32433084990544e-05
};
\addplot [color5, forget plot]
table {%
8 16
8 38.488610056957
};
\addplot [color5, forget plot]
table {%
7.85 1.32433084990544e-05
8.15 1.32433084990544e-05
};
\addplot [color5, forget plot]
table {%
7.85 38.488610056957
8.15 38.488610056957
};
\addplot +[color0,area legend,solid, no marks, fill=white, thick, postaction={pattern=north east lines, pattern color=color0}]
table{
-2.3  6.42375211725632
-1.7  6.42375211725632
-1.7  11.9768267771577
-2.3  11.9768267771577
-2.3  6.42375211725632
};
\path [draw=color0, fill=white, thick, postaction={pattern=north east lines, pattern color=color0}]
(axis cs:3.7,2.79273544534273)
--(axis cs:4.3,2.79273544534273)
--(axis cs:4.3,16)
--(axis cs:3.7,16)
--(axis cs:3.7,2.79273544534273)
--cycle;
\path [draw=color1, fill=white, thick, postaction={pattern=crosshatch, pattern color=color1}]
(axis cs:-1.5,8.00168372464032)
--(axis cs:-0.9,8.00168372464032)
--(axis cs:-0.9,14.2655927728142)
--(axis cs:-1.5,14.2655927728142)
--(axis cs:-1.5,8.00168372464032)
--cycle;
\addplot +[color1,area legend,solid, no marks, fill=white, thick, postaction={pattern=crosshatch, pattern color=color1}]
table{
4.5 1
5.1 1
5.1 12.4497512684613
4.5 12.4497512684613
4.5 1
};
\addplot +[color2,area legend,solid, no marks, fill=white, thick, postaction={pattern=sixpointed stars, pattern color=color2}]
table{
-0.7  8.15664958362689
-0.1  8.15664958362689
-0.1  15.4991117226915
-0.7  15.4991117226915
-0.7  8.15664958362689
};
\path [draw=color2, fill=white, thick, postaction={pattern=sixpointed stars, pattern color=color2}]
(axis cs:5.3,1)
--(axis cs:5.9,1)
--(axis cs:5.9,13.4981681993061)
--(axis cs:5.3,13.4981681993061)
--(axis cs:5.3,1)
--cycle;
\addplot +[color3,area legend,solid, no marks, fill=white, thick, postaction={pattern=bricks, pattern color=color3}]
table{
0.1 7.91783402445769
0.7 7.91783402445769
0.7 14.1303506990947
0.1 14.1303506990947
0.1 7.91783402445769
};
\path [draw=color3, fill=white, thick, postaction={pattern=bricks, pattern color=color3}]
(axis cs:6.1,1)
--(axis cs:6.7,1)
--(axis cs:6.7,13.8835360771815)
--(axis cs:6.1,13.8835360771815)
--(axis cs:6.1,1)
--cycle;
\path [draw=color4, fill=white, thick, postaction={pattern=crosshatch dots, pattern color=color4}]
(axis cs:0.9,6.850641937635)
--(axis cs:1.5,6.850641937635)
--(axis cs:1.5,12.4711901401648)
--(axis cs:0.9,12.4711901401648)
--(axis cs:0.9,6.850641937635)
--cycle;
\addplot +[color4,area legend,solid, no marks, fill=white, thick, postaction={pattern=crosshatch dots, pattern color=color4}]
table{
6.9 1
7.5 1
7.5 15.2751198980687
6.9 15.2751198980687
6.9 1
};
\path [draw=color5, fill=white, thick, postaction={pattern=fivepointed stars, pattern color=color5}]
(axis cs:1.7,6.60707000627882)
--(axis cs:2.3,6.60707000627882)
--(axis cs:2.3,11.4620940090214)
--(axis cs:1.7,11.4620940090214)
--(axis cs:1.7,6.60707000627882)
--cycle;
\addplot +[color5,area legend,solid, no marks, fill=white, thick, postaction={pattern=fivepointed stars, pattern color=color5}]
table{
7.7 1
8.3 1
8.3 16
7.7 16
7.7 1
};
\addplot [thick, black, forget plot]
table {%
-2.3 8.57003055059485
-1.7 8.57003055059485
};
\addplot [thick, black, forget plot]
table {%
3.7 8.04062454786816
4.3 8.04062454786816
};
\addplot [thick, black, forget plot]
table {%
-1.5 10.6374164154873
-0.9 10.6374164154873
};
\addplot [thick, black, forget plot]
table {%
4.5 4
5.1 4
};
\addplot [thick, black, forget plot]
table {%
-0.7 11.1192561428303
-0.1 11.1192561428303
};
\addplot [thick, black, forget plot]
table {%
5.3 4
5.9 4
};
\addplot [thick, black, forget plot]
table {%
0.1 10.4361803707742
0.7 10.4361803707742
};
\addplot [thick, black, forget plot]
table {%
6.1 4
6.7 4
};
\addplot [thick, black, forget plot]
table {%
0.9 9.20804413532703
1.5 9.20804413532703
};
\addplot [thick, black, forget plot]
table {%
6.9 4
7.5 4
};
\addplot [thick, black, forget plot]
table {%
1.7 8.83998792560622
2.3 8.83998792560622
};
\addplot [thick, black, forget plot]
table {%
7.7 4.74255329807471
8.3 4.74255329807471
};
\end{axis}

\end{tikzpicture}}
    \caption{Average \gls{aoi} and error (in terms of state and query \gls{mse}) for different policies in the periodic query scenario.}
    \label{fig:policy_aoi_mse}
\end{figure*}
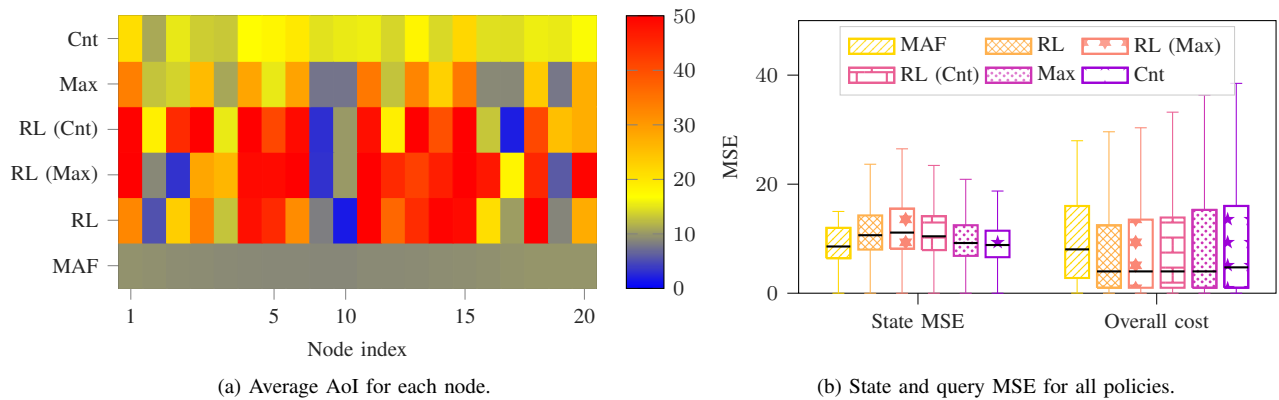

We can first look at the \gls{maf} policy, in Fig.~\ref{fig:color_maf}: the distribution of values is almost symmetrical, and values between -5 and 5 are polled with approximately the same frequency. On the other hand, the index colormap shows a checkerboard pattern, caused by the round robin-like pattern of updates (which is shifted by 2 steps at every cycle, as $N$ is not a multiple of the query period). The \gls{rl} policy has a different pattern: we can note that in even time slots, corresponding to the query instant, the distribution of values is bimodal: sensors whose value is close to 0 are polled very often, as are sensors whose value is very high, between 7 and 12. These two peaks correspond to the two queries: values close to 0 are at the edge of the interval that is relevant for the count range query, while very high values are obviously interesting for the maximum query. The indexes of the sensors that are polled are also much more concentrated: sensors 1, 6, 8, and 17 are polled extremely often, while other sensors are rarely polled: this is due to the nature of the problem, as some sensors are more valuable to answer the queries due to the evolution of the state.

The two single-query \gls{rl} policies, whose choices are represented in Fig.~\ref{fig:color_rlcnt}-\subref*{fig:color_rlmax}, can further shed light on the behavior of the joint policy: we can easily see that some of the nodes that are often polled by \gls{rl} are also polled by its Max and Cnt versions, and that the two peaks in the distribution are close to a superposition of the two peaks of \gls{rl} (Max) and \gls{rl} (Cnt). Finally, we can note that the one-step greedy policies, shown in Fig.~\ref{fig:color_cnt}-\subref*{fig:color_max}, do not have any dependence on the time step, as they are unaware of the query process: the basic features, such as the Cnt policy choosing values clustered around 0 and the Max policy choosing values on the highest end of the range, are maintained, but the policies are inherently noisier than their \gls{rl}-based versions, which can exploit their knowledge of the query process to improve performance at the right moment.

The two plots in Fig.~\ref{fig:policy_aoi_mse} can further clarify the difference between simple \gls{aoi} minimization, \gls{voi} minimization, and query-aware \gls{voi}. Fig.~\ref{fig:aoi_eval} shows the average \gls{aoi} for each sensor for different policies. Naturally, the \gls{maf} policy maintains the minimum \gls{aoi}, with an average only slightly over 10 for all sensors. As the policy tries to minimize the age for all sensors, the average is very similar across all sensors, although not identical (sensors with a higher packet error rate will be polled slightly more often as the poll is repeated after each packet loss). All other policies have a higher \gls{aoi} for some sensors, polling them less often as they have less useful information, and the \gls{rl}-based ones have the highest difference, with some sensors being polled extremely often and others almost never: as information useful for the queries can be reconstructed from the correlation between different sensors and the model of the dynamic system, the \gls{rl} policies rarely poll sensors whose values are less useful or informative for the specific queries they are trained for. The plot in Fig.~\ref{fig:mse_eval}, showing boxplots of the \gls{mse} on the state estimation for each policy clearly shows this: the \gls{rl}-based policies actually have a \emph{higher} \gls{mse} than simple \gls{maf}, as parts of the state are disregarded, but make a significantly smaller error when replying to queries, as the relevant information for the clients is given more importance in the scheduling.

\begin{figure}[t]
    \centering
    \input{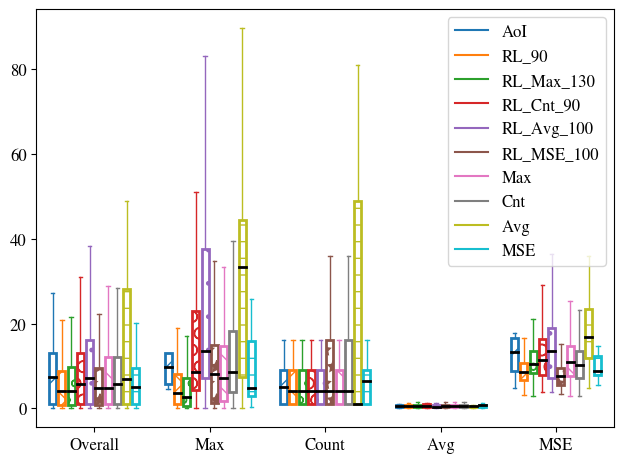}
    \caption{\gls{mse} cost of the different policies for both types of query in the periodic query scenario with $C=4$.}
    \label{fig:periodic_4q}
\end{figure}

Finally, we consider a more complex scenario, with 4 clients asking periodic queries with a period of 12 steps. Along with the maximum and count range queries, the two additional clients make state and sample mean queries, as defined in Sec.~\ref{ssec:queries}. In this case, we also consider \gls{rl}-based and one-step optimal strategies aimed at these queries, and the overall performance in terms of the reward and the specific queries (all of which have the same importance) is shown in Fig.~\ref{fig:periodic_4q}. We note that the average query is easy to respond to, as errors in opposite directions over different components of the state tend to compensate: the \gls{rl} solution outperforms all others in terms of the overall cost (i.e., the \gls{mse} over all queries made by clients), as while each query-specific strategy performs best on its own objective function, the \gls{rl} policy manages to balance different queries, achieving a low error on all of them. We also highlight that legacy \gls{voi} optimization, even when foresighted (i.e., the \gls{rl} (\gls{mse}) strategy in the figure), cannot effectively deal with maximum or count range queries, which depend only on specific parts of the state, as it does not take into account this importance, but statically aims at minimizing the error over all state components indiscriminately.

\subsection{Geometric Query Arrival}
We can consider a second scenario, in which at each step a query of either type is generated with probability $1/6$. The average frequency of queries is the same as for the previous scenario, but instead of a deterministic, periodic sequence, queries follow a memoryless random process with geometrically distributed inter-query times. We remark that queries of both types may arrive to the edge node at the same time, and that in this case, no knowledge is available at the edge node: as the query process is memoryless, knowing the arrival times of past queries provides no information on future query arrivals. In this case, the advantage of a query-aware system is naturally diminished. The time since the last query of each type is still maintained as part of the input to the \gls{rl} algorithm, so as to maintain the same architecture for all cases, but in this case, the \gls{rl} algorithm needs to learn that this information is useless. This case also required more training than the other scenarios we considered.

\begin{figure}[t]
    \centering
    \input{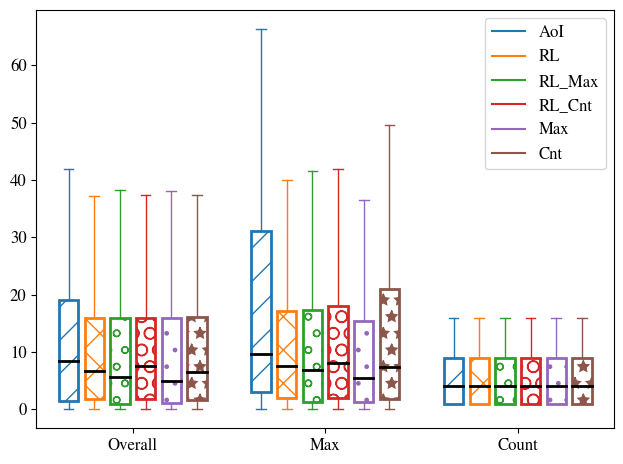}
    \caption{\gls{mse} cost of the different policies for both types of query in the geometric query scenario.}
    \label{fig:Geometric_box}
\end{figure}

Fig.~\ref{fig:Geometric_box} shows the performance boxplots for this scenario: in this case, performance is almost uniform, and all policies have a similar overall cost. The \gls{rl} policy still has a small gain in terms of the overall cost, but it performs worse than the greedy Max policy on the maximum query. Performance on the count range query is almost uniformly good, and all differences between the policies are on the worst-case performance of the maximum query.

\subsection{Mixed Query Arrival}
Finally, we consider a third scenario: in this case, the maximum query follows a memoryless process with a probability $1/6$ of generating a query at each time step, while count range queries are periodically generated every 6 slots. In this case, the policy needs to adapt to the possibility of a maximum query arriving, while also preparing for the foreseen count range queries.

\begin{figure}[t]
    \centering
    \input{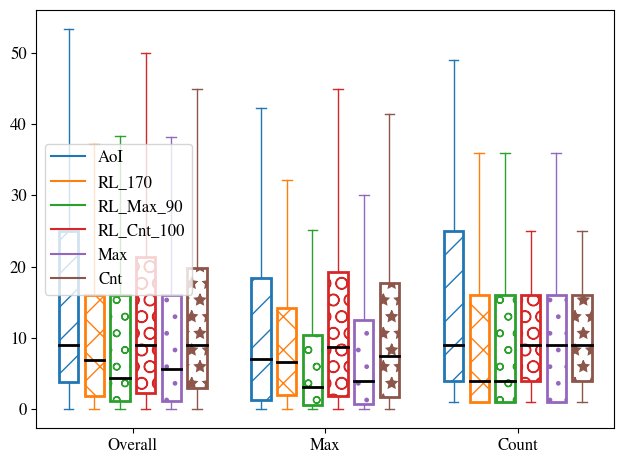}
    \caption{\gls{mse} cost of the different policies for both types of query in the mixed query scenario.}
    \label{fig:mixed_box}
\end{figure}

The performance of the considered policies is shown in the form of boxplots in Fig~\ref{fig:mixed_box}, as for the previous cases. The figure clearly shows that this case is much more complex, and the \gls{rl} policy does not manage to outperform the strategies that are oriented exclusively toward the maximum query. Since the maximum query is entirely unpredictable, the full \gls{rl} policy would need more training to deal with this scenario: the simpler strategy learned by the \gls{rl} (Max) scheme turns out to be better on average, while \gls{rl} (Cnt) performs about as well as \gls{rl}. In most cases, the error on the count range queries tends to be higher for all policies. We note, however, that the \gls{rl} strategy still outperforms all others in terms of worst-case performance, as the 95th percentile whisker is particularly low for the count query, resulting in better overall worst-case performance. A better strategy could be learned with more training, and we note that the complexity of the scenario has a significant impact on the amount of training required, with mixed scenarios with deterministic and stochastic query processes being the most difficult.

By knowing the instants in which count range queries will arrive, the \gls{rl} strategies can limit the worst-case error, although this comes at the cost of a slightly higher worst-case error on the maximum query (which is hard to optimize for, as its arrival process is completely unpredictable). In this case, as in the geometric query arrival scenario, the one-step greedy policy for the maximum query is actually performing almost as well as the \gls{rl} version, as there is no long-term information to be learned on the query process. As for \gls{qaoi}, awareness of the query process is more useful if the latter is deterministic, or at least partially predictable.

\section{Conclusions and Future Work}\label{sec:conc}

In this work, we have presented a framework for query-aware sensor scheduling, in which an edge node needs to choose the most relevant information to respond to external user queries, which may be different functions of the system state. This type of scenario is closely linked to semantic and task-oriented communications in the \gls{iot}, approaching the problem from a different angle: in our system, communications are pull-based, and the bottleneck of the system is medium access rather than rate, so that the solution is semantic, \gls{voi}-based scheduling rather than encoding. Our work shows that query-aware scheduling can lead to profoundly different choices, depending on the specific functions that queries ask for and on the query arrival process for each client, and that \gls{rl}-based strategies can provide a significant advantage in more predictable scenarios, while unpredictable query processes do not provide any useful information to improve scheduling past one-step greedy strategies.

There are several open avenues of research to extend this work, both on the scheduling itself and on the process estimation. Firstly, scheduling is currently limited to a single sensors, and communication is entirely pull-based: a scenario in which multiple sensors can be polled at once, or sensors can transmit urgent information without being polled first, can make scheduling strategies more interesting. Furthermore, extending the problem from simple numeric values to richer types of information such as images or point clouds could prove useful to several applications, such as cooperative driving or robot swarm management, which require the integration of data-heavy information from multiple sources. This also leads to the second line of future work that we are exploring, i.e., the substitution of the Kalman filter with more complex estimators, such as deep networks, which can deal with much more complex functions and system models, and do not require prior knowledge of the system dynamics. Finally, the combination of a control system with the remote estimation would represent another step forward toward a fully task-oriented communication system.

\bibliographystyle{IEEEtran}
\bibliography{bibliography}

\begin{IEEEbiography}[{\includegraphics[width=1in,height=1.25in,clip,keepaspectratio]{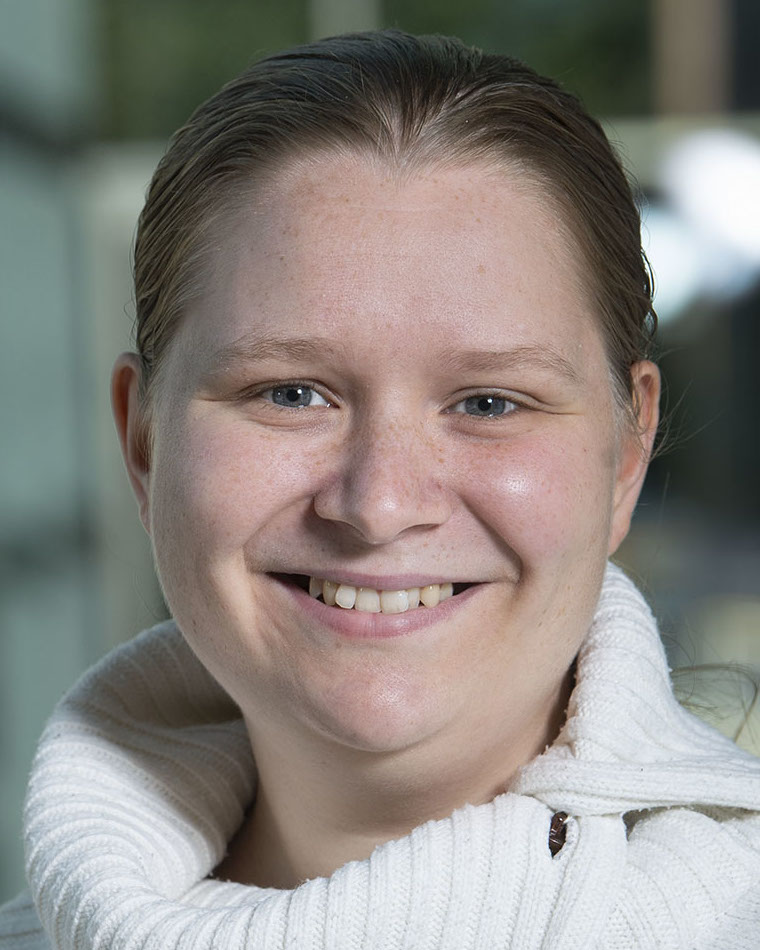}}]
{Josefine Holm} (S'19) received her B.Sc and M.Sc. degrees in mathematical engineering from Aalborg University in 2016 and 2018, respectively. She recently obtained her Ph.D. degree at the Connectivity Section at Aalborg University. Her research interests include wireless communication and IoT networks.
\end{IEEEbiography}

\begin{IEEEbiography}[{\includegraphics[width=1in,height=1.25in,clip,keepaspectratio]{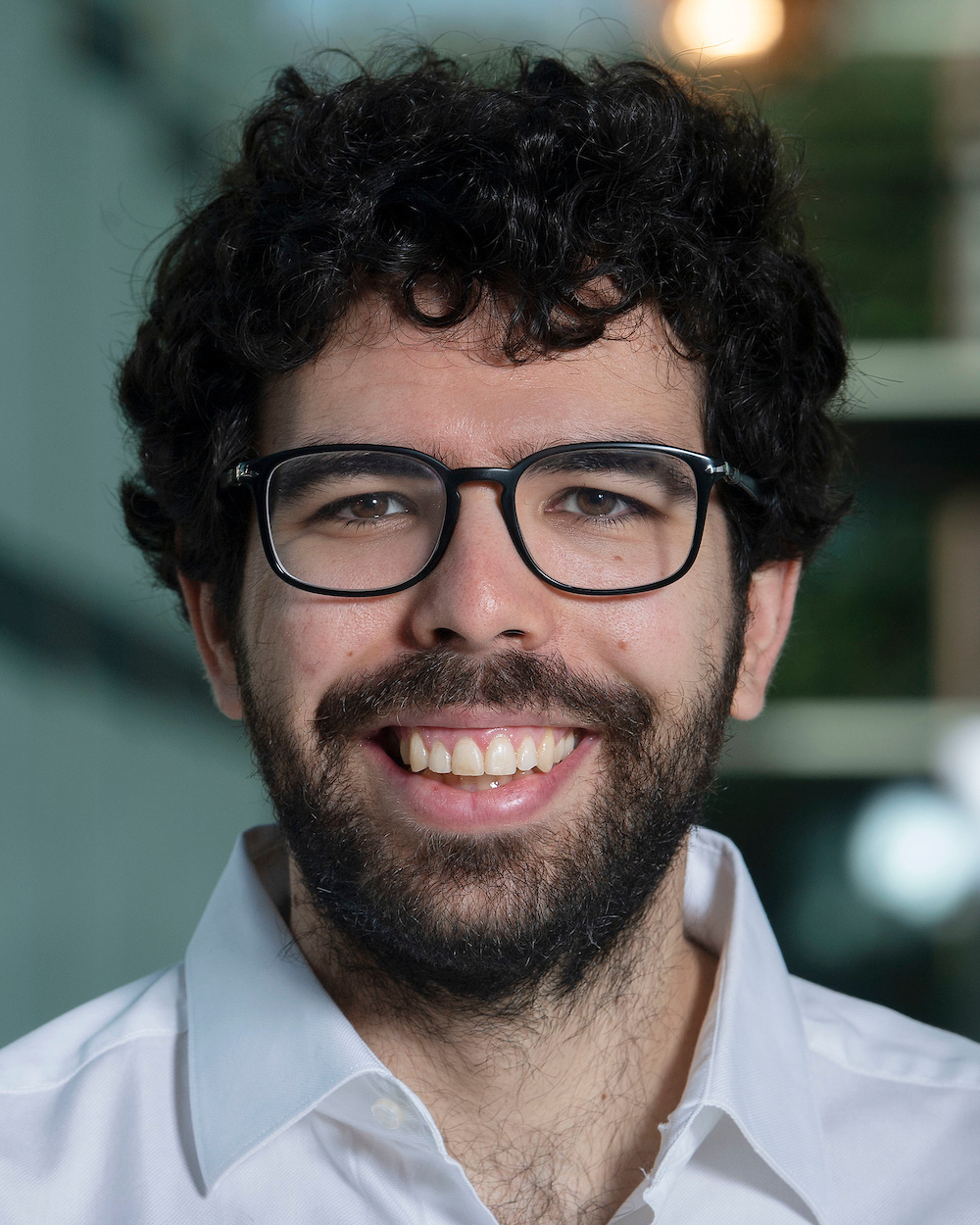}}]
{Federico Chiariotti} (S'15--M'19) is currently an assistant professor at the Department of Information Engineering, University of Padova, Italy, where he also received his Ph.D. in 2019. Between 2020 and 2022, he worked as a post-doctoral researcher and as an assistant professor at the Department of Electronic Systems, Aalborg University, Denmark. He has authored over 60 published papers on semantic communication, Age of Information, Smart Cities, and transport layer protocols. He was a recipient of the Best Paper Award at several conferences, including the IEEE INFOCOM 2020 WCNEE Workshop. His current research interests include network applications of machine learning, transport layer protocols, Smart Cities, bike sharing system optimization, and adaptive video streaming. He is currently an Associate Editor of the IEEE Networking Letters.
\end{IEEEbiography}

\begin{IEEEbiography}[{\includegraphics[width=1in,height=1.25in,clip,keepaspectratio]{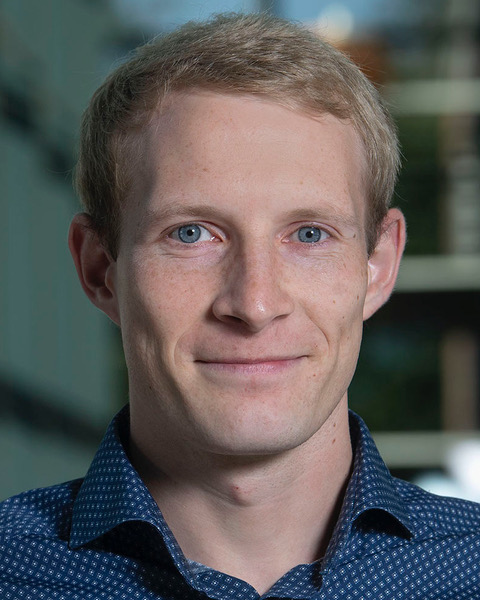}}]
{Anders E. Kal\o r}
(S'17--M'22) received the B.Sc. and M.Sc. degrees in computer engineering in 2015 and 2017, respectively, and the Ph.D. degree in wireless communications in 2022, all from Aalborg University. He is currently a postdoctoral researcher at The University of Hong Kong, supported by an individual International Postdoc grant from the Independent Research Fund Denmark. Concurrently, he is affiliated with the Connectivity section at Aalborg University, Denmark. In 2017, he was a visiting researcher at Bosch, Germany, and in 2020 at King's College London, UK. He was awarded the Spar Nord Foundation Research Award for his Ph.D. project (2023). His current research interests include communication theory and the intersection between wireless communications, machine learning and data mining for IoT.

\end{IEEEbiography}

\begin{IEEEbiography}[{\includegraphics[width=1in,height=1.25in,clip,keepaspectratio]{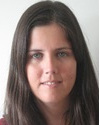}}]
{Beatriz Soret}
(M'11--SM'21) received her M.Sc. and Ph.D. degrees in Telecommunications from the University of Malaga, Spain, in 2002 and 2010, respectively. She is currently a Senior Research Fellow at the Telecommunications Research Institute, University of Malaga, and a part-time Associate Professor at Aalborg University. She has also held industrial positions in Nokia Bell Labs and GomSpace. She received a best paper award in IEEE Globecom 2013 and a Beatriz Galindo senior grant in Spain in 2020. Her current research interests include semantic communications and AoI, LEO satellite communications, and intelligent IoT environments.
\end{IEEEbiography}

\begin{IEEEbiography}[{\includegraphics[width=1in,height=1.25in,clip,keepaspectratio]{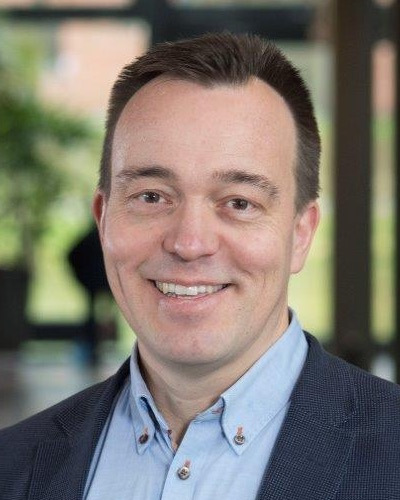}}]
{Torben Bach Pedersen} is a professor
with the Center for Data-Intensive Systems
(Daisy), Aalborg University, Denmark. His
research concerns Predictive, Prescriptive, and Extreme-Scale Data Analytics with Digital Energy as the main application area.
He is an ACM Distinguished Scientist, an IEEE Computer Society Distinguished Contributor, a member of the Danish Academy of
Technical Sciences, and holds an honorary doctorate from TU Dresden.
\end{IEEEbiography}

\begin{IEEEbiography}[{\includegraphics[width=1in,height=1.25in,clip,keepaspectratio]{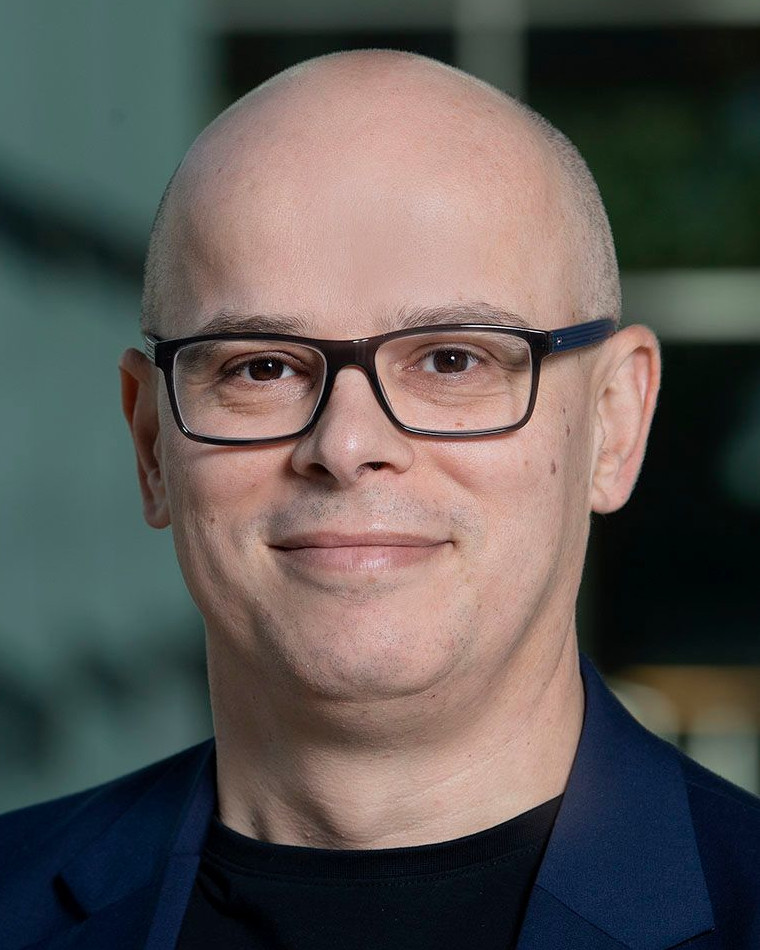}}]
{Petar Popovski} (S'97--A'98--M'04--SM'10--F'16)
is a Professor at Aalborg University, where he heads the section on Connectivity and a Visiting Excellence Chair at the University of Bremen. He received his Dipl.-Ing and M. Sc. degrees in communication engineering from the University of Sts. Cyril and Methodius in Skopje and the Ph.D. degree from Aalborg University in 2005. He is a Fellow of the IEEE. He received an ERC Consolidator Grant (2015), the Danish Elite Researcher award (2016), IEEE Fred W. Ellersick prize (2016), IEEE Stephen O. Rice prize (2018), Technical Achievement Award from the IEEE Technical Committee on Smart Grid Communications (2019), the Danish Telecommunication Prize (2020) and Villum Investigator Grant (2021). He is a Member at Large at the Board of Governors in IEEE Communication Society, Vice-Chair of the IEEE Communication Theory Technical Committee and IEEE TRANSACTIONS ON GREEN COMMUNICATIONS AND NETWORKING. He is currently an Editor-in-Chief of IEEEE JOURNAL ON SELECTED AREAS IN COMMUNICATIONS. Prof. Popovski was the General Chair for IEEE SmartGridComm 2018 and IEEE Communication Theory Workshop 2019. His research interests are in the area of wireless communication and communication theory. He authored the book ``Wireless Connectivity: An Intuitive and Fundamental Guide'', published by Wiley in 2020.
\end{IEEEbiography}
\end{document}